\documentclass[trackchanges]{aastex701}
\usepackage{amsmath,amssymb,amsfonts}
\usepackage{graphicx}
\usepackage{hyperref}
\usepackage{bm}
\usepackage{dcolumn}
\usepackage{listings}
\usepackage{xcolor}
\usepackage{orcidlink}

\begin{document}

\title{A Unified Analytic Framework for Microlensing Caustics: Geode Solutions and Hyper–Catalan Signatures}

\author{Gleb Berloff\,\orcidlink{0009-0009-8457-3589}}
\altaffiliation{Department of Physics and Astronomy}
\affiliation{Department of Physics and Astronomy, University of Warwick, UK}
\email[show]{Gleb.Berloff@warwick.ac.uk}

\author{Natalia G. Berloff\,\orcidlink{0000-0003-2114-4321}}
\altaffiliation{Department of Applied Mathematics and Theoretical Physics}
\affiliation{Department of Applied Mathematics and Theoretical Physics, University of Cambridge, UK}
\email[show]{n.g.berloff@damtp.cam.ac.uk}

\begin{abstract}
We give a preparation‑invariant analytic description of image formation near microlensing caustics. After a local Weierstrass preparation at any multiple image (order $d\ge2$), the lens mapping reduces to a single geode variable $m$ satisfying $m=U\,\varphi(m)$, where $U$ is a prepared source coordinate and $\varphi$ is an image‑side kernel. The coefficients of $m(U)$ obey closed Hyper–Catalan (HC) recurrences, allowing termwise derivatives and truncation control from the characteristic system. We also use the same form for a short HC predictor–corrector: evaluate the series within its certified radius and apply a brief Newton polish near the boundary. We define an HC signature (first nonzero kernel coefficients) and an HC spectrum (branch points and analyticity radius $\rho_U$), which quantify sparsity, stiffness, and safe evaluation domains. The construction covers folds and cusps of any global degree. On a binary fold and cusp, an artificial decic with a resonant unit, and two triple‑lens cusps, HC seeds plus a few Newton steps recover the exact images to machine precision within the certified domain and maintain continuity under continuation. The resulting single‑series templates (with $(\mathrm{Sig}_R,\rho_U)$ metadata) are ready for photometric and astrometric modeling.

\end{abstract}

\keywords{
\uat{Gravitational microlensing}{672} ---
\uat{Binary lens microlensing}{2136} ---
\uat{Astrometry}{80} ---
\uat{Photometry}{1234} ---
\uat{High-resolution microlensing event imaging}{2138}
}

\section{Introduction}\label{sec:intro}

In microlensing, key observables are roots of high‑degree algebraic equations; the hardest regime is near caustics where images appear and merge. Beyond quartics, radicals are unavailable in general, and global root solvers become fragile near folds and cusps. What is needed is a single, differentiable local representation with \emph{a priori} error control across these regions. Analytic caustic expansions improve accuracy and speed (e.g. \citealt{gaudi2002gravitationalI,gaudi2002gravitationalII}); nonetheless, modern pipelines (e.g. \textsc{VBBinaryLensing}, \textsc{MulensModel}; \citep{Bozza2018VBBinaryLensing,PoleskiYee2019MulensModel} remain largely global‑solve based. Here we replace repeated solves by a local series with a certified radius that is uniform across folds and cusps.

The complex-analytic formulation of gravitational lensing by $N$ point masses
places the image coordinate $z=x+\mathrm{i}y$ and the source coordinate
$\zeta=\xi+\mathrm{i}\eta$ in conjugate planes scaled by the Einstein radius of
the total mass.  With lens mass fractions $\{\varepsilon_{j}\}$ satisfying
$\sum_{j=1}^{N}\varepsilon_{j}=1$ and projected lens positions
$s_{j}\in\mathbb{C}$, the mapping is
\begin{equation}
\zeta \;=\; z-\sum_{j=1}^{N}\frac{\varepsilon_{j}}{\overline{z}-\overline{s}_{j}}.
\label{eq:nlens}
\end{equation}
Eliminating $\overline{z}$ recasts this as a single complex polynomial
$P(z;\zeta,\bar \zeta)=0$ whose degree is $N^{2}+1$ for generic configurations; its roots
match the physical images off the caustic discriminant \citep{schneider1986two,asada2009perturbation,rhie2003n}.    Although the eliminant degree grows with $N$, the local micro-geometry is governed by the same two catastrophes (folds and cusps), so from a computational point of view catastrophe-adapted local expansions with certified radii of convergence are strongly preferable to re-solving the full degree-$(N^{2}+1)$ polynomial at every epoch \citep{petters2012singularity,erdl1993classification}.

As the source approaches a caustic, the discriminant of $P$ vanishes and two
(fold) or three (cusp) image roots coalesce.  The Jacobian determinant goes to
zero, point-source magnifications diverge, and global polynomial solvers become
ill-conditioned.  The elimination of $\overline{z}$ also introduces extraneous
roots that must be filtered \citep{asada2009perturbation}, compounding
numerical fragility.  A practical consequence is loss of continuity between
epochs and lack of certified error control precisely where uniform accuracy is
most needed (finite-source photometry, astrometric centroids, and
gradient-based parameter estimation) \citep{gaudi2002gravitationalII,gaudi2002gravitationalI,Witt1994Can,Gould1997Stokes,Bozza2010Microlensing,dominik2004folds, alexandrov2011asymptotic}.
These difficulties motivate preparation-invariant, locally convergent analytic expansions capable of bridging image coalescence and enabling stable fitting.
A locality preserving expansion with
provable convergence and explicit parameter derivatives is therefore required.
For comparison, in the weak-lensing regime, where light deflections induce
small distortions in background galaxy shapes, analytic modeling has long
relied on perturbative basis set frameworks and shear estimators
\citep{refregier2003weak,refregier2003shapeletsI,birrer2015gravitational}.  Microlensing, with its discrete image
multiplicities and catastrophic behavior near caustics, has lacked an
analogous, preparation-invariant analytic framework.

Recent Hyper–Catalan (HC) constructions give closed, purely algebraic series for polynomial roots \citep{wildberger2025hyper, Gessel2016Lagrange,stanton1988recent}. For $p(z)=\sum_{k=0}^{n} c_k z^k$, each root admits an ordinary power series whose coefficients are integer HC multinomials times rational functions of $\{c_k\}$. Evaluation is well‑conditioned under IEEE floating point, derivatives follow termwise, and the associated characteristic system provides a certified convergence radius.
 In
Supplementary Information we provide a Mathematica implementation of
HC series for generic polynomials In Section A, and 
outline the geometric construction of HC coefficients and several practical
advantages of their use in Section B.

We adapt HC series, Lagrange–B\"urmann inversion, and parameter‑dependent Weierstrass preparation to microlensing imaging polynomials. Near any multiple image $(z_\ast,\zeta_\ast)$ of multiplicity $d\!\ge\!2$, centring and rescaling yield a monic local factor and a single “geode’’ variable $m$ so that the mapping takes the Lagrange form $m=U\,\phi(m)$ with a preparation‑invariant kernel $\phi$. All $d$ image branches follow from the one series $m(U)$ and a final single-radical lift.

We encode the local catastrophe by the unit‑normalised kernel $\phi(m)=(1-\sum_{r\ge1}\alpha_r m^r)^{-1}$. The first nonzero $\{\alpha_r\}$ define the HC \emph{signature}; the branch points $(m^\ast,U^\ast)$ of $m=U\,\phi(m)$ and the minimal $|U^\ast|$ define the HC \emph{spectrum} and radius $\rho_U$. These are preparation‑invariant.

Two practical components follow from this geode/HC layer and form the
observables  of the method.  First, near a fold, finite–source
photometry reduces to universal one–dimensional kernels in the scaled normal
offset $s$, with all lens dependence confined to a
handful of scalars read off from the local jets and with truncation tails
certified by the geode/characteristic radius.  This recovers, in a
parameter–explicit form, the fold kernels used to measure source size and limb
darkening
\citep{albrow2001planet,Blackman2020Confirmation,gaudi2002gravitationalI,gaudi2002gravitationalII,Witt1994Can,Gould1997Stokes},
and because the kernels are single–series functions of $s$, their derivatives
with respect to physical parameters remain analytic, enabling gradient–based
fits and Fisher forecasts without numerical differencing.  Second, the
astrometric centroid is described by a single geode series that exhibits the
canonical square–root approach to the fold jump and the standard
cusp–triplet behaviour, again with certified tails, yielding differentiable,
directly usable astrometric approaches relevant to \emph{Gaia}
\citep{belokurov2002astrometric,evans2002microlensing,mcgill2019ongoing},
and in more complex binary–source or binary–lens situations where parallax and
orbital motion affect astrometric paths \citep{Nucita2016BinaryAstrometric}.
Long–baseline surveys such as OGLE now routinely deliver thousands of
microlensing events per year and have enabled population–level inferences on
cold exoplanets, including the finding that wide–orbit ice giants are common
\citep{poleski2021wide}.  Classical fold–crossing phenomenology and
limb–darkening sensitivity in events such as OGLE–1999–BUL–23 and
MACHO–97–BLG–28 can be recovered in the appropriate limits of the same HC
series, while adding certified radii of convergence and an independent of global degree
implementation \citep{albrow2001planet,Blackman2020Confirmation}.  The
upcoming Roman (formerly WFIRST) microlensing survey is expected to detect
$\sim 10^3$ cold planets, including hundreds of sub–Earth–mass planets beyond
1\,AU \citep{Penny2019WFIRST}, making fast, differentiable caustic approaches
particularly valuable. Taken together, these two engines show how the same preparation–invariant
Lagrange form $m=U\,\varphi(m)$, with its HC signature and spectrum can be applied for both photometric and astrometric microlensing.

Our paper is organized as follows.  In
Section~\ref{sec:quintic} we review the binary-lens quintic, its global
discriminant geometry, and image-track evolution away from folds and cusps.  Section~\ref{sec:geode} performs the local geode preparation,
derives the Lagrange form $m=U\,\varphi(m)$, and presents closed HC recurrences
for the coefficients
$M_{n}$ together with preparation–invariant formulas for
the kernel coefficients $\alpha_{r}$ and certified radii from the
characteristic system.  Sections~\ref{sec:ex-fold}--\ref{sec:triple-cusp}  present the binary–lens fold, the binary cusp,
a decic lens with a resonant unit, and a physical triple–lens cusp (presented in
two variants). These illustrate the multiplicity-uniform and independent of the global eliminant degree once locally prepared
character of the geode construction and validate it against exact root tracking,
including certified convergence radii within their analytic domains. Section~\ref{sec:validation} specializes the geode construction to the fold case
($d=2$) and derives preparation-invariant point–source scalings, finite-source
kernels, and astrometric centroids with certified truncation control from local jets.
Section~\ref{sec:HC-classification} develops the preparation–invariant
classification of local microlensing behavior via the Hyper–Catalan signature
and spectrum.  These analytic moduli quantify the sparsity of the geode kernel,
the stiffness of the local mapping, and the certified analyticity radius, thereby
refining the Thom–Arnold fold/cusp taxonomy with data directly relevant to
photometric and astrometric modeling.
Section~\ref{sec:conclusion} summarizes the unified analytic framework, outlines
its degree–agnostic extension to multi–point and multi–plane lenses, and
discusses applications to survey constructions, model comparison, and
gradient-based inference enabled by the differentiable geode/HC representation.
Appendices record implementation details.

\section{Binary--lens mapping, quintic reduction, and analytic kinematics}
\label{sec:quintic}

This section sets up the binary-lens quintic as a compact algebraic testbed and
shows how Hyper–Catalan (HC) series supply not only image positions but also
closed expressions for image velocities and accelerations, together with
differentiable photometric and astrometric observables.  The local
geode/catastrophe machinery in
Sections~\ref{sec:geode}--\ref{sec:HC-classification} remains the main vehicle
for our fold/cusp analysis; the binary quintic here is used primarily for
global kinematics, diagnostics, and validation.

Let $z=x+\mathrm{i}y$ and $\zeta=\xi+\mathrm{i}\eta$ denote image and source
positions (angles) in units of the Einstein radius of the total mass.  Align
the binary along the real axis, place the secondary at $s\in\mathbb{R}$, and
let $\varepsilon=m_{2}/(m_{1}+m_{2})$ be the secondary mass fraction.  The
standard complex lens mapping is \citep{schneider1992properties,chang1979flux,chang1984star,witt1990investigation,an2006chang}

\begin{equation}
\zeta
=
z-\frac{1-\varepsilon}{\overline{z}}-\frac{\varepsilon}{\overline{z}-s}.
\label{eq:lenseq_bin}
\end{equation}
Eliminating $\overline{z}$ with the conjugate of \eqref{eq:lenseq_bin} yields a
single complex polynomial for $z$ whose roots coincide, with algebraic
multiplicity, with the physical images away from the caustic discriminant.  For
a binary lens this eliminant is a quintic
\citep{schneider1986two,witt1990investigation,Witt1994Can}:
\begin{equation}
P(z;\zeta,\overline{\zeta};s,\varepsilon)
=
a_{5}z^{5}+a_{4}z^{4}+a_{3}z^{3}+a_{2}z^{2}+a_{1}z+a_{0}
=0,
\label{eq:quintic_bin}
\end{equation}
whose coefficients are explicit polynomials in $(\zeta,\overline{\zeta})$ and
the parameters $(s,\varepsilon)$:
\begin{equation}
\begin{aligned}
\label{eq:coeffs}
a_{5}&=\overline{\zeta}\,(\overline{\zeta}-s),\\[2pt]
a_{4}&=-s\,\varepsilon+(1+2s^{2})\,\overline{\zeta}+s\,|\zeta|^{2}-2s\,\overline{\zeta}^{2}-|\zeta|^{2}\,\overline{\zeta},\\[2pt]
a_{3}&=s^{2}\,\varepsilon+s\,\zeta-2|\zeta|^{2}-2s^{2}|\zeta|^{2}+s^{2}\,\overline{\zeta}^{2}
+\bigl(-s^{3}+2s(\varepsilon-1)+2s|\zeta|^{2}\bigr)\,\overline{\zeta},\\[2pt]
a_{2}&=s\,\varepsilon-\zeta+s^{2}(\varepsilon-2)\,\zeta
+\bigl(s^{2}-2s^{2}\varepsilon-s^{2}|\zeta|^{2}\bigr)\,\overline{\zeta}
+s^{3}|\zeta|^{2}-2s(\varepsilon-2)\,|\zeta|^{2},\\[2pt]
a_{1}&=s(\varepsilon-1)\Bigl(s\,\varepsilon-(2+s^{2})\,\zeta+2s\,|\zeta|^{2}\Bigr),\\[2pt]
a_{0}&=-s^{2}(\varepsilon-1)^{2}\,\zeta.
\end{aligned}
\end{equation}
The reduction \eqref{eq:quintic_bin} is algebraically equivalent to
\eqref{eq:lenseq_bin} wherever the Jacobian determinant
\begin{equation}
J(z)=1-\left|\frac{\partial\zeta}{\partial\overline{z}}\right|^{2}
=
1-\left|\frac{1-\varepsilon}{\overline{z}^{2}}+\frac{\varepsilon}{(\overline{z}-s)^{2}}\right|^{2}
\label{eq:Jacobian_bin}
\end{equation}
is nonzero.  For generic configurations $P_{z}(z;\zeta,\bar\zeta)$ is
proportional to $J(z)$ up to a nonvanishing factor, so $P_{z}=0$ and $J=0$
pick out the same critical curve.  The quintic $P$ therefore provides a
convenient global vehicle for resultants, discriminants, and the certified
local preparations used later \citep{erdl1993classification,petters2012singularity}.  Signed magnifications are
$\mu_{i}=J(z_{i})^{-1}$.  The critical curve is $J=0$; its image under
\eqref{eq:lenseq_bin} is the caustic
\citep{schneider1992properties,petters2012singularity}.  Along a fold, two
simple roots of $P$ coalesce or are born into a double root; at a cusp, three
simple roots coalesce or are born into a triple root.  Higher multiplicities
arise only on symmetry axes as enforced coincidences.  These are the
$A_{2}/A_{3}$ catastrophes whose local normal forms underlie the geode
construction and the universal kernels employed later \citep{gaudi2002gravitationalI,gaudi2002gravitationalII,petters2012singularity}.  In the linear–fold limit our local formulas reduce to the standard
\(u_\perp^{-1/2}\) magnification law and the \(\tau^{1/2}\) centroid approach, 
providing a direct bridge to the canonical fold procedures
\citep{gaudi2002gravitationalII,gaudi2002gravitationalI}.

\paragraph{Time–dependent quintic and analytic kinematics.}
Let $\zeta(t)$ be any analytic source track and optionally include slow binary
evolution $s(t),\varepsilon(t)$.  Events such as OGLE-2015-BLG-0060, where parallax and binary orbital motion must be modelled jointly \citep{Tsapras2019OGLE2015BLG0060}, illustrate how often 
$
s(t)$ and 
$\varepsilon(t)$ cannot be treated as static. On the source side, xallarap-induced distortions (e.g. OGLE-2015-BLG-0845L, where parallax and source orbital motion combine to allow a lens mass measurement) are increasingly common \citep{Hu2024OGLE0845}. Substituting into \eqref{eq:coeffs} yields
analytic coefficient functions $a_{k}(t)$ and a time-dependent quintic (where we suppressed the dependence on $\zeta, \bar \zeta, \varepsilon$ and $s$)
\begin{equation}
P(z,t)=\sum_{k=0}^{5}a_{k}(t)\,z^{k}=0.
\label{eq:time_dep_quintic}
\end{equation}
Implicit differentiation of \eqref{eq:time_dep_quintic} gives closed forms for
the velocity and acceleration of each image branch $z(t)$:
\begin{equation}
\dot z(t)=-\,\frac{P_{t}(z,t)}{P_{z}(z,t)},\qquad
\ddot z(t)=-\,\frac{P_{tt}+2P_{zt}\dot z+P_{zz}\dot z^{2}}{P_{z}}
\quad\text{evaluated on }P(z,t)=0,
\label{eq:z_dot_ddot}
\end{equation}
where $P_{z}=\partial P/\partial z$, $P_{t}=\partial P/\partial t$, and so on,
with all partials explicit polynomials in $(z,\zeta,\overline{\zeta};s,\varepsilon)$
and their time derivatives.  Because $P_{z}$ vanishes only on the critical
curve, \eqref{eq:z_dot_ddot} is well conditioned away from $J=0$ and matches
the HC–series derivatives obtained by termwise differentiation in the local
charts (Section~\ref{sec:geode}). 

If we want to trace the roots in time, a simple and robust strategy is to
recenter at the current iterate $z_{0}$ and expand
\begin{equation}
P(z_{0}+y,t)=\sum_{j=0}^{5}b_{j}(t;z_{0})\,y^{j},\qquad
b_{j}(t;z_{0})=\sum_{k=j}^{5}\binom{k}{j}a_{k}(t)\,z_{0}^{\,k-j}.
\label{eq:shifted_coeffs}
\end{equation}
Provided $b_{1}\neq 0$, the increment $y$ admits the explicit Hyper–Catalan
series in $(b_{0},b_{2},b_{3},b_{4},b_{5})$:
\begin{equation}
y=\sum_{\substack{m_2,m_3,m_4,m_5\ge 0}}
\frac{(2m_2+3m_3+4m_4+5m_5)!}{(1+m_2+2m_3+3m_4+4m_5)!\,m_2!\,m_3!\,m_4!\,m_5!}\,
\frac{(-b_0)^{1+m_2+2m_3+3m_4+4m_5}\,b_2^{m_2}b_3^{m_3}b_4^{m_4}b_5^{m_5}}
{b_1^{1+2m_2+3m_3+4m_4+5m_5}} .
\label{eq:HC_local}
\end{equation}
This is the specialisation of the general HC root series to a shifted quintic
with coefficients $b_{j}$ evaluated at the base point $z_{0}$.  The multi–index
sum runs over all tuples $(m_{2},m_{3},m_{4},m_{5})$ with total weight
$w = 1+m_{2}+2m_{3}+3m_{4}+4m_{5}$; truncation at a fixed number of terms $K$
provides a rapidly convergent corrector.  A single Newton polish on $P(z,t)=0$
then squares the truncation error.  Because the $b_{j}(t)$ are analytic in $t$,
\eqref{eq:HC_local} differentiates termwise, giving $\dot z$ and $\ddot z$ as
ordinary power series in the same variables; these coincide with
\eqref{eq:z_dot_ddot} away from $J=0$.  In practice we use
\eqref{eq:z_dot_ddot} for cheap global kinematics and the HC–differentiated
series for local, certified error control near catastrophes.

At each step we evaluate the Jacobian from \eqref{eq:Jacobian_bin} and use the
signed magnification to set the step size and to monitor physicality:
\begin{equation}
\Delta t=\frac{\Delta t_{0}}{1+|\mu|},\qquad \mu=J(z_{0})^{-1},\qquad
\mathcal{R}(t)=\bigl|\zeta(z(t))-\zeta(t)\bigr|,
\label{eq:adaptive_dt}
\end{equation}
with a user–chosen base step $\Delta t_{0}$ and a cap on $|\mu|$ near the
discriminant.  The residual $\mathcal{R}(t)$, computed from
\eqref{eq:lenseq_bin}, guards against branch slips.  When $b_{1}\to 0$ the HC
series is replaced by the geode variable $m$ of multiplicity $d=2$ (fold) or
$d=3$ (cusp), as developed in Section~\ref{sec:geode}.

\paragraph{Short tracks and validation geometry.}
For a concrete geometry used for illustration we reuse a wide binary with
$(s,\varepsilon)=(2,1/4)$ and two short tracks: a fold–crossing track
\begin{equation}
\zeta(t)=0.30-0.05\,(t-1)\,\mathrm{i},\qquad t\in[0,2],
\label{eq:fold_track}
\end{equation}
and a cusp–passage track
\begin{equation}
\zeta(t)=t,\qquad t\in[0,1].
\label{eq:cusp_track}
\end{equation}
Along \eqref{eq:fold_track} a pair is created at the entry fold, evolves inside
the critical curve, and  a pair of  images annihilates at the exit fold.  Along
\eqref{eq:cusp_track} a triple root resolves into three branches with the
expected $1/3$ Puiseux exponents.  These tracks serve as running examples after
the local geode preparation is introduced in Section~\ref{sec:geode} and will be
reused in the  cusp example of
Section~\ref{sec:binary-cusp}.

\begin{figure}[h!]
\centering
\includegraphics[width=.9\linewidth]{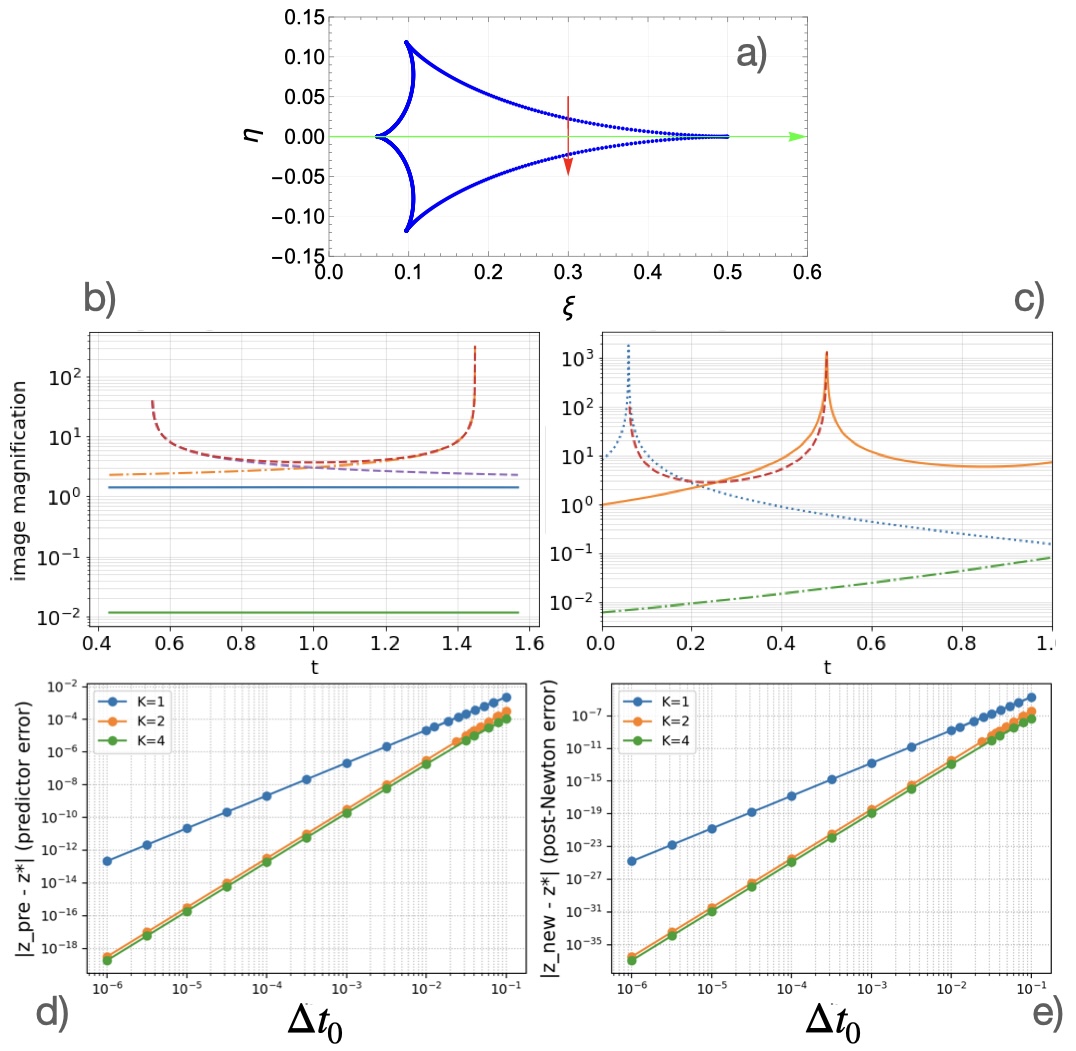}
\caption{ Binary‑lens geometry at $\epsilon=0.25$, $s=2$. (a) Caustics with a fold‑crossing track $\zeta(t)=0.30-0.05(t-1)i$ (red) and a cusp‑passage track $\zeta(t)=t$ (green). (b–c) Per‑image magnifications along each track. (d–e) Predictor and post‑Newton errors versus base step $\Delta t_0$ for HC truncations $K\in\{1,2,4\}$, showing the expected HC scaling and a quadratic reduction after one Newton step.
}
\label{fig:wideBinary}
\end{figure}

Figure~\ref{fig:wideBinary}(a) shows the caustics and the two source tracks used
in \eqref{eq:fold_track}--\eqref{eq:cusp_track}; panels
\ref{fig:wideBinary}b--\ref{fig:wideBinary}c report the per–image magnifications
$|\mu|$ versus $t$ along the two tracks; and panels
\ref{fig:wideBinary}d--\ref{fig:wideBinary}e quantify predictor and post–Newton
errors as functions of the base step $\Delta t$ for HC truncation orders
$K\in\{1,2,4\}$.  The observed error scaling foreshadows the certified
convergence results proved in Section~\ref{sec:geode}. 

\section{Geode construction, Lagrange normalisation, and Hyper--Catalan solution}
\label{sec:geode}

In this section we develop the core geode construction: a
preparation–invariant, single–series parametrisation of all image branches that
bifurcate at a multiple root of multiplicity $d\ge 2$ of the eliminated
single–plane lens polynomial
$
P(z;\zeta,\bar\zeta)=0,
$
with $P$ holomorphic in $z$ and real–analytic in $(\zeta,\bar\zeta)$.  All
examples in Sections~\ref{sec:ex-fold}--\ref{sec:triple-cusp} are concrete
instances of the construction introduced here.
Let $(z_{*},\zeta_{*})$ be such a multiple image, i.e.
$
\partial_{z}^{k}P(z_{*};\zeta_{*},\bar\zeta_{*})=0,$ $(0\le k<d),$
$\partial_{z}^{d}P(z_{*};\zeta_{*},\bar\zeta_{*})\neq0.
$

\paragraph{Shift and normalisation of local coordinates.}
Translate and rescale so that the multiple image sits at the origin and the
pure $z$–jet of order $d$ is monic:
\begin{equation}
z=z_{*}+ \frac{w}{\lambda},\qquad
\zeta=\zeta_{*}+v,\qquad
\bar\zeta=\bar\zeta_{*}+\bar v,
\label{eq:scaling}
\end{equation}
with
\begin{equation}
\lambda=\Bigl[\partial_{z}^{d}P(z_{*};\zeta_{*},\bar\zeta_{*})/d!\Bigr]^{1/d}.
\label{eq:lambda}
\end{equation}
The normalised local polynomial
\begin{equation}
P_{\rm loc}(w;v,\bar v)
=
P\!\bigl(z_{*}+w/\lambda;\,\zeta_{*}+v,\,\bar\zeta_{*}+\bar v\bigr)
\label{eq:Pnorm}
\end{equation}
is centred at $(w,v,\bar v)=(0,0,0)$ with
$\partial_{w}^{d}P_{\rm loc}(0;0,0)/d!=1$, fixing the local geometric scale.
From now on, all derivatives and coefficients in this section are taken in the
centred, normalised coordinates of \eqref{eq:Pnorm}; for brevity we keep the
notation $P_{\rm loc}$ to emphasise this localisation.

\textit{The steps in our local construction are at $(0,0,0)$.} 
(1) Parameter‑dependent Weierstrass preparation: $P_\mathrm{loc}(w;v,\bar v)=U(t;v,\bar v)W(t;v,\bar v)$ with $t=w-w_c(v,\bar v)$. 
(2) Geode variable $m=t^d$. 
(3) Cyclotomic product and square‑free reduction in $m$. 
(4) Lagrange normalisation $m=U\,\varphi(m)$ with prepared source $U=\Lambda(v,\bar v)$. 
(5) HC coefficients from $\varphi$ and a certified radius from the characteristic system. 
(6) Lift back to images $z=z_\ast+(w_c+\omega m^{1/d})/\lambda$ and, if desired, polish by Newton on $W=0$ (or $P=0$). Next,  we will explain these steps in detail.

\paragraph{(i) Weierstrass preparation and centre.}
Let $t=w-w_{c}(v,\bar v)$, where the \emph{Weierstrass centre} $w_{c}$ is the
unique analytic germ with $w_{c}(0,0)=0$ that cancels the $t^{d-1}$ term of the
monic factor $W(t;v,\bar v)$.  By parameter–dependent Weierstrass preparation
\citep{weierstrass1879einige}, there exist an analytic unit
$\mathcal{U}(t;v,\bar v)$ with $\mathcal{U}(0;0,0)\neq0$ and a monic polynomial
$W(t;v,\bar v)$ of degree $d$ such that
\begin{equation}
P_{\rm loc}\bigl(w_{c}(v,\bar v)+t;v,\bar v\bigr)
=
\mathcal{U}(t;v,\bar v)\,W(t;v,\bar v)=0,
\label{eq:WPT}
\end{equation}
with
\begin{equation}
W(t;v,\bar v)=t^{d}+a_{d-2}(v,\bar v)\,t^{d-2}+\cdots+a_{0}(v,\bar v),
\qquad [t^{d-1}]\,W\equiv0,\qquad a_{k}(0,0)=0.
\end{equation}
The Weierstrass centre $w_{c}(v,\bar v)$ is obtained by enforcing
$\partial_{w}^{\,d-1}P_{\rm loc}(w_{c};v,\bar v)=0$.  We always divide the local factor by the pure $t^d$ coefficient and choose the Weierstrass centre $w_c$ so that $[t^{d-1}]W\equiv 0$; hence $[t^d]W\equiv 1$ and $[t^{d-1}]W\equiv 0$ by construction. In code, this should be asserted at every evaluation point. 
 Weierstrass preparation is essential because it isolates a monic
polynomial factor $W$ in the local variable $t$ from a nonvanishing analytic
unit $\mathcal U$.  Only after this factorisation does the substitution
$t=\omega^{j}m^{1/d}$ produce a single–valued analytic function of the geode
variable $m$; without it, fractional powers and nonpolynomial terms in $w$
would obstruct the construction.

\paragraph{(ii–iii) Geode, cyclotomic product, and square–free reduction.}
Introduce the geode variable $m=t^{d}$ and set $\omega=\mathrm{e}^{2\pi i/d}$.
Form the $d$–fold cyclotomic product
\begin{equation}
R(m;v,\bar v)
=
\prod_{j=0}^{d-1}
P_{\rm loc}\!\bigl(w_{c}(v,\bar v)+\omega^{j}m^{1/d};\,v,\bar v\bigr)
=
\Biggl[\prod_{j=0}^{d-1}\mathcal{U}\!\bigl(\omega^{j}m^{1/d};v,\bar v\bigr)\Biggr]
\Biggl[\prod_{j=0}^{d-1}W\!\bigl(\omega^{j}m^{1/d};v,\bar v\bigr)\Biggr],
\label{eq:cyclo}
\end{equation}
The first bracket is an analytic unit in $m$; the second is a polynomial in $m$
of degree $d$ whose zeros are the geode images $m=t^{d}$ of the roots of
$W(t;v,\bar v)=0$. We also define the Weierstrass cyclotomic product
\begin{equation}
    R_W(m;v,\bar v)=\prod_{j=0}^{d-1}W\!\bigl(\omega^{j}m^{1/d};v,\bar v\bigr).
    \label{eq:RW}
\end{equation}

The substitution $t\mapsto t^{d}$ may introduce a parameter–uniform multiplicity in
$m$ which we now remove.  Let
\begin{equation}
\widetilde R(m;v,\bar v)
=
\frac{R(m;v,\bar v)}{\gcd_{m}\bigl(R,\partial_{m}R\bigr)}, \quad 
\widetilde R_W(m;v,\bar v)
=
\frac{R_W(m;v,\bar v)}{\gcd_{m}\bigl(R_W,\partial_{m}R_W\bigr)}
\label{eq:Rsf}
\end{equation}
with the gcd taken in the coefficients
analytic in $(v,\bar v)$.  For folds ($d=2$), the cyclotomic product has the
form $R\propto\bigl(m+p_{0}(v,\bar v)\bigr)^{2}$ and hence
\begin{equation}  
\widetilde R_W=m+p_{0}(v,\bar v)
\label{eq:2d}
\end{equation}
after dividing out the common factor.  For
$d\ge3$, when the local cubic (or higher) factor $W$ has simple roots
(i.e. the discriminant $\Delta_{m}(R)$ does not vanish identically near
$(v,\bar v)=(0,0)$), $R$ and $\partial_{m}R$ are generically coprime, so
$\gcd_{m}(R,\partial_{m}R)=1$ and $\widetilde R\equiv R$ up to an analytic unit.
However, special parameter values can produce an \emph{accidental} extra factor
of $m$ in $R(m;v,\bar v)$ (so that $R$ and $\partial_{m}R$ share an $m^{k}$
factor).  In such cases the square–free reduction \eqref{eq:Rsf} cancels this
extra multiplicity.  An explicit example of
this accidental multiplicity and its removal appears in the near-tangential
triple-lens cusp of Section~\ref{sec:triple-cusp}.

\paragraph{(iv) Lagrange normalisation and prepared source coordinate.}
 Expand the square–free Weierstrass cyclotomic  product as
\begin{equation}
\widetilde R_W(m;v,\bar v) = A(v,\bar v)\,m - B(v,\bar v) + m^{2}\Psi(m;v,\bar v)=0,
\label{eq:RABC}
\end{equation}
with $A=\partial_{m}\widetilde R_W(0;v,\bar v)$ and $B=-\widetilde R_W(0;v,\bar v)$,
and define the prepared source coordinate
\begin{equation}
U=\Lambda(v,\bar v)=\frac{B(v,\bar v)}{A(v,\bar v)}.
\label{eq:U}
\end{equation}
Dividing (\ref{eq:RABC}) by $A$ gives the geometric Lagrange form
\begin{equation}
m = U\,\varphi_{\rm geo}(m,U),
\qquad
\varphi_{\rm geo}(m,U)=\Bigl[1+\frac{m\,\Psi(m;v,\bar v)}{A(v,\bar v)}\Bigr]^{-1}.
\label{eq:phigeo}
\end{equation}

If $d=2$ (fold), the square–free scalar is
$\widetilde R_W(m;v,\bar v)=m+p_0(v,\bar v)$, so $A\equiv1$, $B=-p_0$ and
$U=\Lambda(v,\bar v)=-p_0(v,\bar v)$. Hence the geode equation is strict:
$
m=U,$ $ \varphi\equiv1.
$

\paragraph{(v) Preparation–invariant kernel and Hyper–Catalan series}
Evaluate the square–free scalar at the base point $(v,\bar v)=(0,0)$:
\[
\widetilde R(m;0,0)=\sum_{k\ge d} r_k\,m^{k},\qquad r_d\neq0.
\]
For $d\ge3$ set
\begin{equation}
D(m)=\frac{\widetilde R(m;0,0)}{r_d\,m^{d}}
=1+\sum_{r\ge1}d_r\,m^{r},
\qquad d_r=[m^{r}]\bigl(D(m)-1\bigr),
\label{eq:D-from-R}
\end{equation}
and define the \emph{preparation–invariant kernel}
\begin{equation}
\varphi(m)=\frac{1}{D(m)}
=\frac{r_d\,m^{d}}{\widetilde R(m;0,0)},
\qquad \varphi(0)=1.
\label{eq:phi-from-R}
\end{equation}
If the cyclotomic product contributes a nonconstant pure $m$-unit, then $\widetilde R(m;0,0)=r_d\,m^{d}\varphi(m)^{-1};$
this retains any resonances (e.g. the decic example of Section \ref{sec:decic-resonant} with $\varphi(m)=(1+m^{7})^{-1}$).
Together with the prepared source $U=\Lambda(v,\bar v)$ this yields the \emph{HC Lagrange form}
\begin{equation}
m=U\,\varphi(m),\qquad \varphi(0)=1,
\label{eq:Lmaster}
\end{equation}
in which $U$ carries the \emph{source-side} transport while $\varphi$ encodes the \emph{image-side} geometry at the base point.
All preparation–dependent constants enter through $\Lambda$, not through $\varphi$.

The unique analytic germ solving \eqref{eq:Lmaster} is the HC series
\begin{equation}
m(U)=\sum_{n\ge1}M_n\,U^{n},\qquad
M_n=\frac{1}{n}\,[w^{\,n-1}]\,\varphi(w)^{n}.
\end{equation}
Writing the kernel in denominator form
\begin{equation}
\varphi(m)=\Bigl(1-\sum_{r\ge1}\alpha_r\,m^{r}\Bigr)^{-1}
\label{eq:phi-den}
\end{equation}
gives the explicit Hyper–Catalan coefficients
\begin{equation}
M_n=\frac{1}{n}\!
\sum_{\{k_r\}\,:\,\sum r k_r=n-1}
\frac{\bigl(n+\sum k_r-1\bigr)!}{(n-1)!\,\prod_r k_r!}
\prod_{r\ge1}\alpha_r^{\,k_r}.
\label{eq:LB}
\end{equation}

\paragraph{Characteristic system and analyticity radius.}
Branch points of $m(U)$ are critical points of \eqref{eq:Lmaster}:
\begin{equation}
m^{*}=U^{*}\,\varphi(m^{*}),\qquad 1=U^{*}\,\varphi'(m^{*}).
\label{eq:char}
\end{equation}
The analyticity radius about $U=0$ is
\begin{equation}
\rho_{U}=\min\bigl\{|U^{*}|:\ \exists\,m^{*}\ \text{satisfying \eqref{eq:char}}\bigr\}.
\label{eq:rhoU}
\end{equation}
Equivalently, in the Weierstrass coordinates of \eqref{eq:WPT} the branch points
correspond to singular solutions of the local polynomial factor
$
W(t^{*};v^{*},\bar v^{*})=0,$ $
\partial_{t}W(t^{*};v^{*},\bar v^{*})=0,
$
with $m^{*}=(t^{*})^{d}$ and $U^{*}=\Lambda(v^{*},\bar v^{*})$.  The minimal
modulus $|U^{*}|$ among such pairs determines $\rho_{U}$.

\paragraph{Certified tails via a majorant kernel.}
Define the nonnegative majorant kernel
\begin{equation}
\widehat\varphi(m)
=
\Bigl(1-\sum_{r\ge1}|\alpha_{r}|\,m^{r}\Bigr)^{-1},
\label{eq:phihat}
\end{equation}
and let $\widehat m(U)=\sum_{n\ge1}\widehat M_{n}U^{n}$ solve
$\widehat m=U\,\widehat\varphi(\widehat m)$ with $\widehat m(0)=0$.  Then
$\widehat M_{n}\ge0$ and $|M_{n}|\le \widehat M_{n}$ for all $n$, and for any
$|U|<\widehat\rho_{U}$ (the analyticity radius induced by $\widehat\varphi$)
\begin{equation}
\Bigl|\,m(U)-\sum_{n=1}^{N}M_{n}U^{n}\Bigr|
\;\le\;
\widehat m(|U|)
\;-\;
\sum_{n=1}^{N}\widehat M_{n}\,|U|^{\,n}.
\label{eq:tail}
\end{equation}
The right–hand side is explicit once $D(m)$ (and hence the $\alpha_{r}$) are
known.

\paragraph{(vi) Reconstruction and practical algorithm.}
Given a source offset $(v,\bar v)$ near $(0,0)$, reconstruction proceeds as follows:
\begin{enumerate}\itemsep0.25em
\item[(1)] Compute the prepared coordinate
$U=\Lambda(v,\bar v)=B(v,\bar v)/A(v,\bar v)$ from \eqref{eq:RABC}.

\item[(2)] Form the preparation–invariant kernel $\varphi(m)$ from
\eqref{eq:phi-from-R} (for $d\ge 3$). For folds ($d=2$) use the linear kernel
obtained directly from \eqref{eq:2d}.

\item[(3)] Evaluate the geode series $m(U)=\sum_{n\ge1}M_{n}U^{n}$ via \eqref{eq:LB},
truncating at order $K$ and using \eqref{eq:tail} to certify the truncation error.

\item[(4)] Lift to $d$ geode seeds $t^{(0)}_{k}=\omega^{k} m^{1/d}$ with $\omega^{d}=1$.
For $d=3$ the monic Weierstrass factor takes the depressed form
\begin{equation}
W(t;v,\bar v)=t^3 + p(v,\bar v)\,t + q(v,\bar v),\qquad [t^3]W\equiv 1,\ [t^2]W\equiv 0.
\end{equation}
When the skewness $S=\lvert p\rvert/\lvert m\rvert^{2/3}$ is not small, the three
naive lifts $t^{(0)}_{k}=\omega^{k} m^{1/3}$ can lie in the same Newton basin and
polish to a single branch.  A robust fix is to use the first Cardano
correction,
\begin{equation}
t^{(1)}_k \;=\; \omega^k m^{1/3}\;-\;\frac{p}{3\,\omega^k m^{1/3}},\qquad k=0,1,2,
\label{eq:anisotropic-seed}
\end{equation}
which is exactly the one-step Newton update for $t^{3}+pt+q=0$ linearized about
$t=\omega^{k} m^{1/3}$.

\item[(5)] (Optional) Refine each seed by Newton iteration on the centred Weierstrass
factor $W(t;v,\bar v)=0$ to obtain the exact local roots $t_{k}$.

\item[(6)] Map back to the physical image plane via
\[
z=z_{*}+\frac{w_{c}(v,\bar v)+t_{k}}{\lambda},
\]
and, if desired, perform the final Newton polish on the full eliminated equation
$P(z;\zeta,\bar\zeta)=0$ as a global consistency check.
\end{enumerate}

Thus the HC series provides analytic seeds, while the exact images are recovered
by a short, stable Newton solve on $W=0$ (or $P$). The construction cleanly separates
preparation–dependent data $(W,w_{c},\Lambda)$ from preparation–invariant
objects $(\varphi,\{\alpha_{r}\},\{M_{n}\},\rho_{U})$, with $\varphi$ computed solely from
the square–free scalar $\widetilde R(m;0,0)$.

\paragraph{Remark on roots collapse.}
The geode lift $t\sim\omega\,m^{1/3}$ is only the leading term of the local expansion for each branch of a cusp. 
If the geometric coefficients $\{p,q\}$ in the monic cubic $W=t^3+pt+q$ are not small relative to $m^{2/3}$ and $m$, two of the true roots may sit well outside the circle $|t|=|m|^{1/3}$.
In that regime the naive seeds $t^{(0)}_k=\omega^k m^{1/3}$ may lie in the same Newton basin (near the small–$|t|$ root) and hence “collapse’’ under Newton iteration.
Adding the $O(p/m^{1/3})$ correction in \eqref{eq:anisotropic-seed} restores the correct basins while preserving the preparation–invariant Lagrange map $m=U\,\varphi(m)$.

\paragraph{Universality and link to examples.} The construction applies to any analytic lens polynomial with a finite‑multiplicity image: all local behaviour is captured by $m=U\,\varphi(m)$. The intrinsic data are $\{\alpha_r\}$ and $\rho_U$; preparation‑dependent choices appear only through $\Lambda$. Sections~\ref{sec:ex-fold}–\ref{sec:triple-cusp} illustrate folds, binary cusps, a resonant decic, and triple‑lens cusps.
 Notations and symbols used are presented in the Supplementary Information in Section C
 for a quick reference.

\section{ Artificial quintic with a fold}
\label{sec:ex-fold}

This example implements the full geode construction in the simplest fold case
($d=2$), where the preparation–invariant kernel is trivial,
$\varphi\equiv 1$.  It therefore isolates, in a transparent setting,
centering and local normalisation, geode elimination, Lagrange
normalisation, certified analyticity radius, and Newton polish.

Consider
\begin{equation}\label{eq:P_fold_start}
P(z,\zeta)=z^{5}+z^{2}+\zeta\,(z^{3}+z+1).
\end{equation}
At $(z,\zeta)=(0,0)$ there is a multiple image of multiplicity $d=2$ since
$P(z,0)=z^{2}(z^{3}+1)$.  We work at the base point
$(z_{*},\zeta_{*})=(0,0)$ and scale $\lambda$ as in \eqref{eq:lambda}; here
$\lambda=1$ since $\partial_{z}^{2}P(0;0)/2!=1$.

\paragraph{(i) Local normalisation and centre $z_{c}(\zeta)$.}
Set $z=z_{c}(\zeta)+t$ and choose $z_{c}(\zeta)$ so that the linear term in $t$
vanishes:
$
[t^{1}]\,P\bigl(z_{c}(\zeta)+t,\zeta\bigr)=0.
$
Expanding in $\zeta$ around $\zeta=0$ gives the centred image position
\begin{equation}\label{eq:center_series}
z_{c}(\zeta)=-\tfrac{1}{2}\,\zeta-\tfrac{3}{8}\,\zeta^{3}-\tfrac{5}{32}\,\zeta^{4}
+O(\zeta^{5}).
\end{equation}
For a quadratic Weierstrass factor $W=t^{2}+p_{0}(\zeta)$ one can express
$z_{c}(\zeta)$ and $p_{0}(\zeta)$ in terms of $P(0,\zeta)$,
$P_{z}(0,\zeta)$, and $P_{zz}(0,\zeta)$ via
\begin{equation}\label{eq:zc-p0-leading}
z_{c}(\zeta)
=-\,\frac{P_{z}(0,\zeta)}{P_{zz}(0,\zeta)}+O(\zeta^{2}),
\qquad
p_{0}(\zeta)
=\frac{P(0,\zeta)}{P_{zz}(0,\zeta)}
-\frac{P_{z}(0,\zeta)^{2}}{4\,P_{zz}(0,\zeta)^{2}}
+O(\zeta^{2}),
\end{equation}
which, in this example, reduce to the explicit series
\eqref{eq:center_series} and \eqref{eq:W-fold-example} below.

\paragraph{(ii–iii) Monic Weierstrass factor, geode, and square–free
eliminant.}
Dividing by the $t^{2}$–coefficient yields the monic Weierstrass factor
\begin{equation}\label{eq:W-fold-example}
W(t,\zeta)=t^{2}+p_{0}(\zeta),
\qquad
p_{0}(\zeta)=-\zeta+\tfrac{1}{4}\zeta^{2}-\tfrac{3}{2}\zeta^{3}
-\tfrac{3}{4}\zeta^{4}+O(\zeta^{5}).
\end{equation}
Introduce the geode $m=t^{2}$ to symmetrise the two branches.  The $d=2$
cyclotomic product
\begin{equation}
R(m,\zeta)=\prod_{j=0}^{1}
P\bigl(z_{c}(\zeta)+\omega^{j} m^{1/2},\zeta\bigr),
\qquad \omega^2=1,
\end{equation}
factorises as
\begin{equation}\label{eq:R-factor}
R(m,\zeta)=\mathcal{U}(m,\zeta)\,\bigl(m+p_{0}(\zeta)\bigr)^{2},
\qquad
\mathcal{U}(m,\zeta)=\mathcal{U}\!\bigl(\sqrt{m},\zeta\bigr)\,
\mathcal{U}\!\bigl(-\sqrt{m},\zeta\bigr),\ \ \mathcal{U}(0,0)\neq 0,
\end{equation}
where $\mathcal{U}$ is the analytic Weierstrass unit.
For folds, the pure $m$–part of $\mathcal{U}$ at the base point is constant,
so it does not enter the kernel.  Removing the parameter–uniform multiplicity
introduced by $m=t^{2}$ gives the square–free eliminant
\begin{equation}\label{eq:simple_fold}
\widetilde R_W(m,\zeta)=m+p_{0}(\zeta)=0.
\end{equation}

\paragraph{(iv) Canonical fold Lagrange form and kernel $\varphi\equiv1$.}
Equation \eqref{eq:simple_fold} is already in Lagrange form
\begin{equation}\label{eq:Lagrange_fold}
m=U\,\varphi(m),
\qquad
U=\Lambda(\zeta)=-p_{0}(\zeta),
\qquad
\varphi(m)\equiv 1,\ \ \varphi(0)=1.
\end{equation}
Thus the preparation–invariant kernel is strict, with no resonant unit:
$\alpha_{r}\equiv0$ and the HC series collapses to $m(U)=U$ exactly.  In terms
of the source coordinate, $m(\zeta)=-p_{0}(\zeta)$, and the two local images
are $t_{\pm}=\pm m^{1/2}$.  In the classification language of
Section~\ref{sec:HC-classification}, this corresponds to an HC signature with
multiplicity $d=2$ and no nonzero kernel coefficients, i.e.\ a canonical fold
with trivial kernel.

\paragraph{(v) Seeds, Newton polish, and lift.}
Use geode seeds $t^{(0)}_{\pm}=\pm m^{1/2}$ and refine them by Newton’s method
on the well–conditioned quadratic $W=t^{2}+p_{0}$:
\begin{equation}\label{eq:newton_fold}
t\ \leftarrow\ t-\frac{W(t,\zeta)}{\partial_{t}W(t,\zeta)}
= t-\frac{t^{2}+p_{0}(\zeta)}{2t}.
\end{equation}
In implementation, one guards this update for very small $|t|$ (or directly
uses the exact quadratic root $t=\pm\sqrt{-p_{0}}$) to avoid division by $2t$.
Undoing the shift (and scale) gives the images
\begin{equation}\label{eq:back_to_z}
z(\zeta)=z_{*}+z_{c}(\zeta)+t(\zeta),\qquad (\lambda=1).
\end{equation}
Thus the HC series supplies analytic seeds and the exact images are obtained by
a single, stable Newton polish on $W=0$.

\paragraph{(vi) Certified analyticity radius.}
Because the Weierstrass unit is nonvanishing, the systems $(P,P_{z})=(0,0)$
and $(W,W_{t})=(0,0)$ are equivalent; we use $(P,P_{z})$ for convenience.
From
\begin{equation}\label{eq:zeta-of-t}
P_{z}(t,\zeta)=5t^{4}+2t+\zeta(3t^{2}+1)=0
\ \Rightarrow\
\zeta(t)=-\frac{5t^{4}+2t}{3t^{2}+1},
\end{equation}
eliminating $\zeta$ from $P=0$ and $P_{z}=0$ (e.g.\ via the resultant in
$\zeta$) yields the critical equation
\begin{equation}\label{eq:crit_poly}
2+t+4t^{3}+4t^{4}+2t^{6}=0.
\end{equation}
Its nearest roots $t^{\ast}$ map by \eqref{eq:zeta-of-t} to
\begin{equation}\label{eq:zetastar}
\zeta^{\ast}\approx -0.0230696\pm 0.2700834\,i,\qquad
\rho_{\zeta}=\min|\zeta^{\ast}|\approx 0.271.
\end{equation}
Hence the local series $m(\zeta)$ (and thus $t(\zeta)$) converge for
$|\zeta|<\rho_{\zeta}$.  Since $U=\Lambda(\zeta)=-p_{0}(\zeta)$ and
$\Lambda'(0)\neq0$, the corresponding analyticity radius in the source
coordinate satisfies $\rho_{U}\sim|\Lambda'(0)|\,\rho_{\zeta}$ near the base
point, in agreement with the general characteristic–radius discussion of
Section~\ref{sec:geode}.  Cauchy majorants give explicit tail bounds at any
$r\in(|\zeta|,\rho_{\zeta})$, and one Newton update on $W$ squares the geode
seed error.

\paragraph{Numerical check at \(\zeta=0.01\).}
From \eqref{eq:center_series} and \eqref{eq:W-fold-example}:
\[
z_{c}(0.01)=-5.000377\times 10^{-3},\qquad
p_{0}(0.01)=9.976508\times 10^{-3},\qquad
t^{(0)}_{\pm}=\pm i\,9.988247\times 10^{-2}.
\]
Thus
\[
z_{\pm}^{(0)}=z_{c}(0.01)+t^{(0)}_{\pm}
=-5.000377\times 10^{-3}\ \pm\ i\,9.988247\times 10^{-2}.
\]
One Newton step using \eqref{eq:newton_fold} already reduces the residuals
$|P(z_{\pm},0.01)|$ to the $10^{-3}$ level; in practice a second Newton step
drives the residuals to machine precision at $|\zeta|=0.01\ll\rho_{\zeta}$.

This example realises the general scheme with $d=2$, $m(U)=U=-p_{0}(\zeta)$, and
trivial kernel $\varphi\equiv1$. 
Figure~\ref{fig:examples}(a) shows the two image branches as
$\zeta$ is varied along the real axis.  Solid curves (geode + Newton on $W$)
and dashed curves (direct root solving) agree to within numerical precision
throughout $|\zeta|\lesssim 0.6$.  Within the certified radius
$|\zeta|<\rho_{\zeta}\approx 0.271$ this agreement is guaranteed by the
characteristic–radius analysis, and the numerical performance remains excellent
even somewhat beyond this domain.

\section{ Binary--lens cusp (physical quintic)}
\label{sec:binary-cusp}

We now apply the geode construction to a physical binary-lens cusp.  For the
binary with $(s,\varepsilon)=(2,1/4)$ the single–plane eliminant
$P(z;\zeta,\bar\zeta)$ has a triple image ($d=3$) at
$
(\zeta_{*},z_{*})=\bigl(0.0603074,\,-0.8793852\bigr),
$
corresponding to a cusp on the central caustic.  For notational simplicity, we will assume the source track is along the the real axis, so $\zeta=\bar \zeta$. Following the construction of
Section~\ref{sec:geode}, we translate and scale
\begin{equation}\label{eq:bin-pullback}
z=z_{*}+\frac{w}{\lambda},\qquad
\zeta=\zeta_{*}+u,\qquad
P_{\rm loc}(w;u)=P(z_{*}+w/\lambda;\,\zeta_{*}+u),
\end{equation}
with $\lambda$ chosen so that the pure $w$–jet is monic (Eq.~(\ref{eq:lambda})), so
$
\lambda=\Bigl[\partial_{z}^{3}P(z_{*};\zeta_{*})/3!\Bigr]^{1/3}
=-0.8398545975895852.
$
This sets $\partial_{w}^{3}P_{\rm loc}(0;0)/3!=1$ and fixes the local scale.

\paragraph{(i) Weierstrass centre and monic factor.}
Let $t=w-w_{c}(u)$; the Weierstrass centre $w_{c}(u)$ is the unique analytic
germ with $\partial_{w}^{2}P_{\rm loc}\!\bigl(w_{c}(u);u\bigr)=0$ and
$w_{c}(0)=0$.  Solving order by order in $u$ gives
\begin{equation}\label{eq:bin-wc}
w_{c}(u)=
-35.2886 u + 518.743 u^2 - 12003.3 u^3 + 43246.5 u^4
+O(u^5).
\end{equation}
By parameter–dependent Weierstrass preparation,
\begin{equation}\label{eq:bin-W}
P_{\rm loc}\!\bigl(w_{c}(u)+t;u\bigr)=\mathcal U(t,u)\,W(t,u),\qquad
W(t,u)=t^{3}+p(u)\,t+q(u),\quad p(0)=q(0)=0,
\end{equation}
with $\mathcal U(0,0)\neq0$ and a depressed ($[t^{2}]W=0$) monic cubic $W$.
This is the $d=3$ specialisation of the general preparation
\eqref{eq:WPT}.

\paragraph{(ii–iii) Geode, cyclotomic product, and square–free scalar.}
Introduce the cusp geode $m=t^{3}$ and $\omega=\mathrm e^{2\pi i/3}$.  The
cyclotomic product removes fractional powers and is analytic in $m$,
\begin{equation}\label{eq:bin-cyclo}
\widetilde R_W(m;u)=R_W(m;u)=\prod_{j=0}^{2}
W\!\bigl(\omega^{j}m^{1/3};\,u\bigr).
\end{equation}

\paragraph{(iv) Prepared source and geometric Lagrange form.}
Expand the square–free scalar associated with the local cubic factor $W$ as
\begin{equation}\label{eq:RABC-binary}
\widetilde R_W(m;u)
=
A(u)\,m - B(u) + m^{2}\Psi(m;u),\qquad
A(u)=\partial_{m}\widetilde R_W(0;u),\;
B(u)=-\widetilde R_W(0;u),
\end{equation}
and define the prepared source
\begin{equation}\label{eq:U-Lambda-binary}
U=\Lambda(u)=\frac{B(u)}{A(u)}
= 42394.0757\,u^{3} + O(u^{4}).
\end{equation}
This $U(u)$ parameterises the exact geometric transport for our chosen
preparation: dividing \eqref{eq:RABC-binary} by $A(u)$ gives the geometric
Lagrange form
$
m = U\,\varphi_{\rm geo}(m,U),$ $
\varphi_{\rm geo}(m,U)
=\Bigl(1+\tfrac{m\,\Psi(m;u)}{A(u)}\Bigr)^{-1},
$
with $u$ regarded as a function of $U$ via \eqref{eq:U-Lambda-binary}.

\paragraph{(v) Preparation–invariant kernel $\varphi$, HC series, and radius.}
Independently of this geometric transport, the preparation–invariant kernel
used for HC seeds is obtained from the full imaging polynomial $P$ by forming
its cyclotomic product in the geode variable $m$ and then taking the square–free
scalar at the base point $(u=0)$.  For the present binary–lens cusp this yields
the pure $m$–jet
\[
\widetilde R(m;0)=m^{3}\bigl[\,1-\alpha_{1}m-\alpha_{2}m^{2}\,\bigr],
\]
so the preparation–invariant kernel is
$
\varphi(m)=(1-\alpha_{1}m-\alpha_{2}m^{2})^{-1},$ $ \varphi(0)=1,
$
with
$
\alpha_{1}=-0.4203353,$ $
\alpha_{2}=-0.0219406,$ $ \alpha_{r\ge3}=0.
$
In the HC classification of Section~\ref{sec:HC-classification}, this cusp has
signature $\mathrm{Sig}_{R}=(3;\alpha_{1},\alpha_{2})$.

For this binary example the cyclotomic unit from the full $P$ does not carry
any nontrivial $m$–dependence at the base point, so $\widetilde R_W$
and $\widetilde R$ agree up to an analytic unit and the geometric kernel
$\varphi_{\rm geo}$ and preparation–invariant kernel $\varphi$ coincide at
$(m,U)=(0,0)$.  In higher–degree cases (e.g. the decic cusp with a resonant
unit in Section~\ref{sec:decic-resonant}) the cyclotomic product of the full $P$ does
contribute a nontrivial pure $m$–unit, and the preparation–invariant kernel
$\varphi$ must be computed from this full square–free scalar rather than from
the cubic factor alone.

With the quadratic kernel $\varphi(m)$, only $\alpha_{1}$ and $\alpha_{2}$
enter the HC recurrences.  Specialising \eqref{eq:LB} to this case gives
\[
M_{n}=\frac{1}{n}
\sum_{k_{2}=0}^{\lfloor (n-1)/2\rfloor}
\frac{\bigl(2(n-1)-k_{2}\bigr)!}{(n-1)!\,\bigl((n-1)-2k_{2}\bigr)!\,k_{2}!}\,
\alpha_{1}^{\,n-1-2k_{2}}\alpha_{2}^{\,k_{2}},\qquad n\ge1,
\]
and the geode series
\[
m(U)=\sum_{n\ge1}M_{n}U^{n}.
\]

For truncation control we use the nonnegative majorant kernel
$\widehat\varphi(m)=(1-|\alpha_{1}|m-|\alpha_{2}|m^{2})^{-1}$, which induces
$\widehat m(U)=\sum \widehat M_{n}U^{n}$ with $\widehat M_{n}\ge0$ and
$|M_{n}|\le\widehat M_{n}$.  The majorant characteristic equation
$3|\alpha_{2}|m^{2}+2|\alpha_{1}|m-1=0$ has a positive root
$\widehat m^{*}$, and the corresponding certified radius
$
\widehat\rho_{U}
=\widehat m^{*}\bigl(1-|\alpha_{1}|\,\widehat m^{*}
-|\alpha_{2}|(\widehat m^{*})^{2}\bigr)
=0.5622011
$
is a direct specialisation of the general characteristic–radius formula
\eqref{eq:rhoU}.

\paragraph{(vi) Reconstruction at \(u=0.01\) and Newton polish.}
To reconstruct images at a specific source position we use the practical
algorithm of Section~\ref{sec:geode}.  For $u=0.01$ the exact Weierstrass
centre, obtained by solving $\partial_{w}^{2}P_{\rm loc}(w;u)=0$, is
$
w_{c}(0.01)=-0.3139924.
$
For comparison, the truncated series \eqref{eq:bin-wc} gives
$w_{c}(0.01)\approx -0.312582$.  At this centre define
$f(t;u)=P_{\rm loc}(w_{c}(u)+t;u)$ and evaluate
$
f_{0}=f(0;u),\quad f_{1}=\partial_{t}f(0;u),\quad
f_{3}=\tfrac{1}{6}\partial_{t}^{3}f(0;u)
$
at $u=0.01$, obtaining
$
f_{0}=-0.076677,$ $
f_{1}=0.1752943,$ $
f_{3}=0.4313251.
$
Hence the prepared source
$$
U=\frac{B(u)}{A(u)}
=-\frac{f_{0}^{3}}{3 f_{0}^{2} f_{3} + f_{1}^{3}}
=0.0346934.
$$
(Using the full $U(u)$ expression \eqref{eq:U-Lambda-binary} yields
$U\approx 0.0342434$.)

Solving the cubic
$
-\alpha_2 m^3 - \alpha_1 m^2 + m - U = 0
$
for \(m\) at this \(U\) gives
$
m = 0.0342009.
$
Lifting to geode seeds $t_{k}$ with Cardano correction Eq.~(\ref{eq:anisotropic-seed}) gives
\begin{equation}
    t_k=\{0.310598 -0.477264 \, i,-0.406215+0.193705 \, i,0.0956164 +0.283559 \, i\}.
\end{equation}
Mapping back to the physical image plane by
$z=z_{*}+(w_{c}(u)+t_{k})/\lambda$ and performing six Newton polish steps gives the three images at $u=0.01$ to 
 machine accuracy.

This binary–lens cusp realises the geode construction for $d=3$ with a nontrivial
kernel
$\varphi(m)=(1-\alpha_{1}m-\alpha_{2}m^{2})^{-1}$ and HC signature
$\mathrm{Sig}_{R}=(3;\alpha_{1},\alpha_{2})$.  The prepared source
$U(u)$ plays the role of the local control parameter; the HC
series $m(U)$ provides analytic seeds with certified radius
$|U|\le\widehat\rho_{U}$; Newton on the centred cubic $W(t,u)$ yields the exact
images; and the lift
$z=z_{*}+\bigl(w_{c}(u)+\omega\,m^{1/3}\bigr)/\lambda$ reconstructs the three
branches.  The full evolution of these images as the source moves along the
real axis is shown in Figure~2(b).  Solid curves (HC + Newton on $W$) and
dashed curves (direct numerical solutions of $P$) are indistinguishable within
the certified domain $|U(u)|\le\widehat\rho_{U}$, including the neighbourhood
of the cusp where the three branches coalesce. This cusp illustrates the universal behavior predicted by catastrophe theory and confirmed in observational modeling of quasar microlensing caustic crossings \citep{shalyapin2002nature}.
Our framework generalizes such approaches to degree–agnostic analytic representations. Together with
Example~\ref{sec:ex-fold}, this demonstrates how the same local geode
framework accommodates both folds and cusps in physical binary lenses.

\section{ Decic cusp with a resonant unit}
\label{sec:decic-resonant}

Our third example is a cusp whose eliminated polynomial is of degree ten and
carries a nontrivial \emph{resonant unit} in the geode variable.  It provides a
clean test of the HC construction in a higher–degree setting with a nontrivial
kernel.

Consider
\begin{equation}\label{eq:decic-P}
P(z,\zeta)=z^{10}+z^{3}+\zeta\,(1+z+z^{7}+z^{8}).
\end{equation}
At $\zeta=0$ there is a triple image at $z_{*}=0$ ($d=3$), since
$P(z,0)=z^{3}(1+z^{7})$.  With $z=t$ and unit scaling $\lambda=1$ the local
Weierstrass preparation takes the simple form
\begin{equation}\label{eq:decic-WP}
P_{\rm loc}(t,\zeta)=\mathcal U(t,\zeta)\,W(t,\zeta),
\qquad
\mathcal U(t,\zeta)=1+t^{7},\qquad
W(t,\zeta)=t^{3}+\zeta(1+t),
\end{equation}
so $z_{c}\equiv 0$ and $W$ is already a depressed cubic.

\paragraph{(ii–iii) Geode, cyclotomic product, and square–free scalar.}
Set the cusp geode $m=t^{3}$ and $\omega=e^{2\pi i/3}$.  The three–fold
cyclotomic product is
\begin{equation}\label{eq:decic-R}
R(m,\zeta)
=\prod_{j=0}^{2}P(\omega^{j}m^{1/3},\zeta)
=\Big[(m+\zeta)^{3}+\zeta^{3}m\Big]\,
(1+m^{7}),
\end{equation}
where the bracket arises from the cubic factor $W$ and the unit $1+m^{7}$
comes from the cyclotomic product of $\mathcal U$.  Near $(m,\zeta)=(0,0)$
one has $\gcd_{m}(R,\partial_{m}R)=1$, so the square–free scalar is
$
\widetilde R(m,\zeta)\equiv R(m,\zeta)
$
up to a nonzero constant factor.

\paragraph{(iv) Geometric transport.}
Write the geometric bracket in \eqref{eq:decic-R} as
\begin{equation}\label{eq:decic-ABpsi}
(m+\zeta)^{3}+\zeta^{3}m
= A(\zeta)\,m - B(\zeta) + m^{2}\Psi(m,\zeta),
\quad
A(\zeta)=\zeta^{2}(3+\zeta),\; B(\zeta)=-\zeta^{3},\; \Psi(m,\zeta)=m+3\zeta.
\end{equation}
The prepared source coordinate is
\begin{equation}\label{eq:decic-U}
U=\Lambda(\zeta)=\frac{B(\zeta)}{A(\zeta)}=-\frac{\zeta}{3+\zeta},
\end{equation}
which behaves as $U\sim -\zeta/3$ near $\zeta=0$.  Then the \emph{exact
geometric transport} identity obtained from $A\,m-B+m^{2}\Psi=0$ is
\begin{equation}\label{eq:geo-transport}
U
= m\Bigl(1+\frac{m\,\Psi(m,\zeta)}{A(\zeta)}\Bigr)
= m+\frac{m^{2}\,(m+3\zeta)}{\zeta^{2}(3+\zeta)}
\quad\Longleftrightarrow\quad
m=U\,\varphi_{\rm geo}(m,\zeta),\ \ 
\varphi_{\rm geo}(m,\zeta)=\Bigl(1+\tfrac{m\,\Psi}{A}\Bigr)^{-1}.
\end{equation}
The kernel $\varphi_{\rm geo}$ encodes the exact $m$–$U$ relation for this
decic, but depends explicitly on $\zeta$ and is not preparation–invariant.

\paragraph{(v) Preparation–invariant kernel \(\varphi\), resonant unit,
and HC seeds.}
Independently of the geometric transport \eqref{eq:geo-transport}, the
\emph{preparation–invariant} kernel that drives the HC seeds is obtained at the
base point from the square–free scalar:
\begin{equation}\label{eq:phi-from-R_decic}
\widetilde R(m;0)=m^{3}(1+m^{7})
\quad\Rightarrow\quad
r_{3}=[m^{3}]\widetilde R=1,\qquad
\varphi(m)=\frac{r_{3}m^{3}}{\widetilde R(m;0)}
=\frac{1}{1+m^{7}}.
\end{equation}
Thus $\varphi$ records the \emph{resonant unit} $1+m^{7}$ coming from the
cyclotomic product; it is \emph{not} the full geometric kernel
$\varphi_{\rm geo}$ in \eqref{eq:geo-transport}.  In the HC classification of
Section~\ref{sec:HC-classification}, this cusp has signature
\begin{equation}
\mathrm{Sig}_{R}=(3;\alpha_{7}),\qquad \alpha_{7}=-1,\ \ \alpha_{r\neq7}=0,
\end{equation}
reflecting the seven–fold resonance in the kernel.

Let $m(U)=\sum_{n\ge1}M_{n}U^{n}$ solve $m=U\,\varphi(m)$ with
$\varphi$ from \eqref{eq:phi-from-R_decic}.  By Lagrange–B\"urmann,
\[
M_{n}=\frac{1}{n}[w^{\,n-1}]\varphi(w)^{n}
=\frac{1}{n}[w^{\,n-1}](1+w^{7})^{-n},
\]
so $M_{n}$ vanishes unless $n\equiv 1\pmod 7$.  Explicitly,
\begin{equation}\label{eq:decic-Mn}
M_{n}=
\begin{cases}
(-1)^{k}\dfrac{1}{7k+1}\binom{8k}{k}, & n=7k+1,\\[6pt]
0, & n\not\equiv 1\pmod 7,
\end{cases}
\qquad
m(U)=U-U^{8}+8U^{15}-92U^{22}+1240U^{29}-\cdots.
\end{equation}
This HC series furnishes analytic \emph{seeds}; the exact geometric transport is
restored using \eqref{eq:geo-transport} after lifting to $t$ and polishing on
$W=0$.

\paragraph{Characteristic system and seed analyticity radius.}
Branch points of the seed map $m(U)$ satisfy
\[
m^{*}=U^{*}\,\varphi(m^{*}),\qquad 1=U^{*}\,\varphi'(m^{*})
\ \Longleftrightarrow\ 
\varphi(m^{*})-m^{*}\varphi'(m^{*})=0.
\]

For \(\varphi(m) = (1 + m^7)^{-1}\) we have
\[
\varphi'(m) = -\frac{7 m^6}{(1 + m^7)^2},\qquad
\Delta(m) = \varphi(m) - m \varphi'(m)
= \frac{1 + 8 m^7}{(1 + m^7)^2}.
\]
Thus branch points satisfy \(1 + 8 m^7 = 0\), i.e.
\(m^7 = -1/8\), so \(|m_\ast| = (1/8)^{1/7}\).
The corresponding control values are
\[
U_\ast = \frac{m_\ast}{\varphi(m_\ast)} = m_\ast(1 + m_\ast^7)
= m_\ast\Bigl(1 - \frac{1}{8}\Bigr) = \frac{7}{8}m_\ast,
\]
and the analyticity radius of the HC seed series is therefore
$
\rho_U = \min |U_\ast| = \frac{7}{8}\left(\frac{1}{8}\right)^{1/7}
\simeq 0.6506.
$
In practice the usable domain is further constrained by the source transport
$U=\Lambda(\zeta)=-\zeta/(3+\zeta)$ and by proximity to the actual cusp in the
$\zeta$–plane (see Figure~2(c)).

\begin{figure}[h!]
\centering
\includegraphics[width=\textwidth]{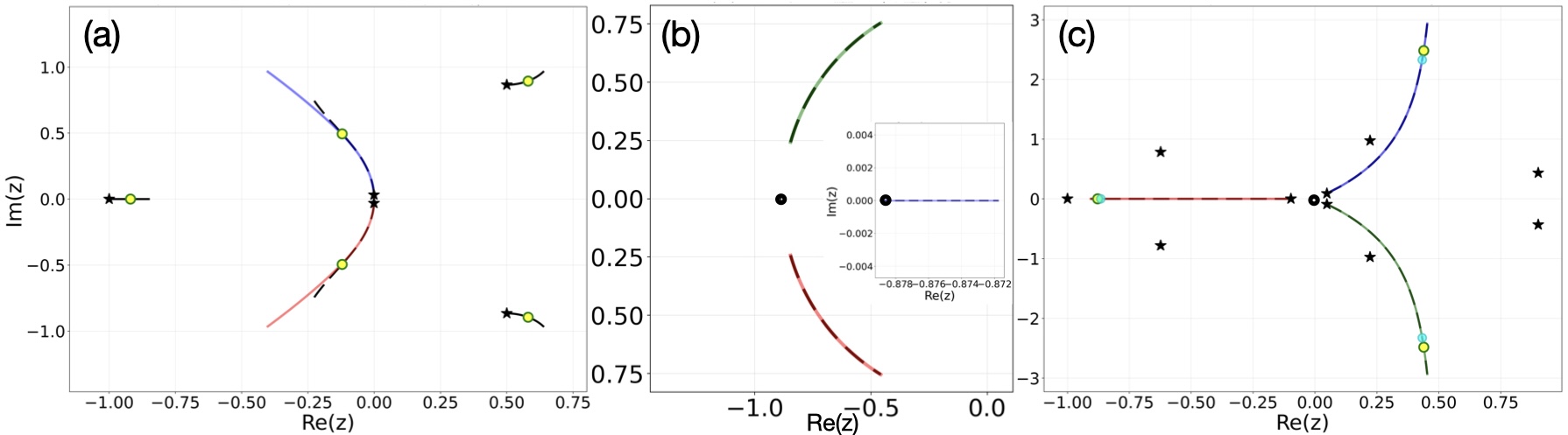}
\caption{
{\bf Root trajectories for three polynomial systems} as the parameter 
$\zeta$ (or $u=\zeta-\zeta_{*}$) varies.
(a) \textbf{Quintic fold} 
given by Eq.~\ref{eq:P_fold_start}.
Black stars mark the roots at $\zeta\approx0$.  
Solid coloured curves (blue/red) show geode approximations valid up to $\zeta=0.6$.  
Black dashed lines indicate numerically exact roots extending to $\zeta=0.8$.  
Yellow–green circles mark the convergence radius $\rho_{\zeta}\approx0.271$.
(b) \textbf{Binary–lens cusp} considered in Section \ref{sec:binary-cusp} emanating from $(z_{*},\zeta_{*})$.  
Three image branches (coloured solid: geode; black dashed: exact) 
are tracked from the cusp singularity as the parameter $u$ increases 
until $|U(u)|$ reaches the certified radius $\widehat{\rho}_{U}$.  
The black filled circle marks the cusp point.  
Inset: zoomed view of the local structure near the cusp with the third image trajectory too short to be visible on the main plot.
(c) \textbf{Decic polynomial} (given by Eq.~\ref{eq:decic-P}).
Black stars correspond to all ten roots at $\zeta\approx0$.  
Coloured solid curves show the HC  series 
combined with Newton polishing for the triple image.  
Black dashed lines show the   exact roots trajectories emanating from the cusp.  
Cyan circles mark the safe radius ($0.95\,\rho_{U}$) where continuation begins, 
while yellow–green circles denote the convergence radius $\rho_{U}$. HC seeds from the unit–aware kernel $\varphi(m) = (1 + m^7)^{-1}$ are used only for
\(|U| < 0.95\,\rho_U\) with \(\rho_U \simeq 0.651\); beyond this limit a predictor–corrector
continuation uses the previous polished roots as seeds.
All three panels demonstrate geode–based root tracking with high accuracy 
within their certified radii of convergence.
}
\label{fig:examples}
\end{figure}

\paragraph{Reconstruction (practical route).}
Reconstruction in this example follows the general recipe:
\begin{enumerate}\itemsep2pt
\item Given $\zeta$, compute the prepared source
$U=\Lambda(\zeta)=-\zeta/(3+\zeta)$.
\item Evaluate the HC seed $m(U)$ from \eqref{eq:decic-Mn}, truncating at
$|U|<\rho_{U}$ and using the sparsity pattern $n=7k+1$.
\item Lift to cubic seeds $t^{(0)}_{k}=\omega^{k}m(U)^{1/3}$, $k=0,1,2$.
\item Enforce the exact relation by Newton iteration on the centred factor
$W(t,\zeta)=t^{3}+\zeta(1+t)$:
\[
t\ \leftarrow\ t-\frac{W(t,\zeta)}{3t^{2}+\zeta}.
\]
\item Output images $z=t$ (since $z_{c}\equiv 0$).  Error transport from $m$ to
$t$ obeys
\[
|\delta t^{(0)}|\le \tfrac{1}{3}|m|^{-2/3}|\delta m|,
\]
and Newton converges quadratically whenever $3t^{2}+\zeta\neq 0$.
\end{enumerate}
Here the resonant unit $1+m^7$ affects only the seed kernel $\varphi$; the $\zeta$‑dependence enters via the analytic transport and the cubic $W$, on which Newton gives quadratic convergence.
Figure~\ref{fig:examples}(c)  shows the ten roots of the
decic as $\zeta$ varies along the real axis.  The coloured solid curves
(HC seeds + Newton polish on $W$) reproduce the three physical branches
emanating from the cusp and lie directly atop the numerically exact root tracks
(black dashed curves), up to the analyticity boundary set by $\rho_{U}$.  This
demonstrates that the HC geode framework remains stable, accurate, and
degree-agnostic even in the presence of a nontrivial resonant kernel.


\begin{figure}[h!]
\centering
\includegraphics[width=0.9
\textwidth]{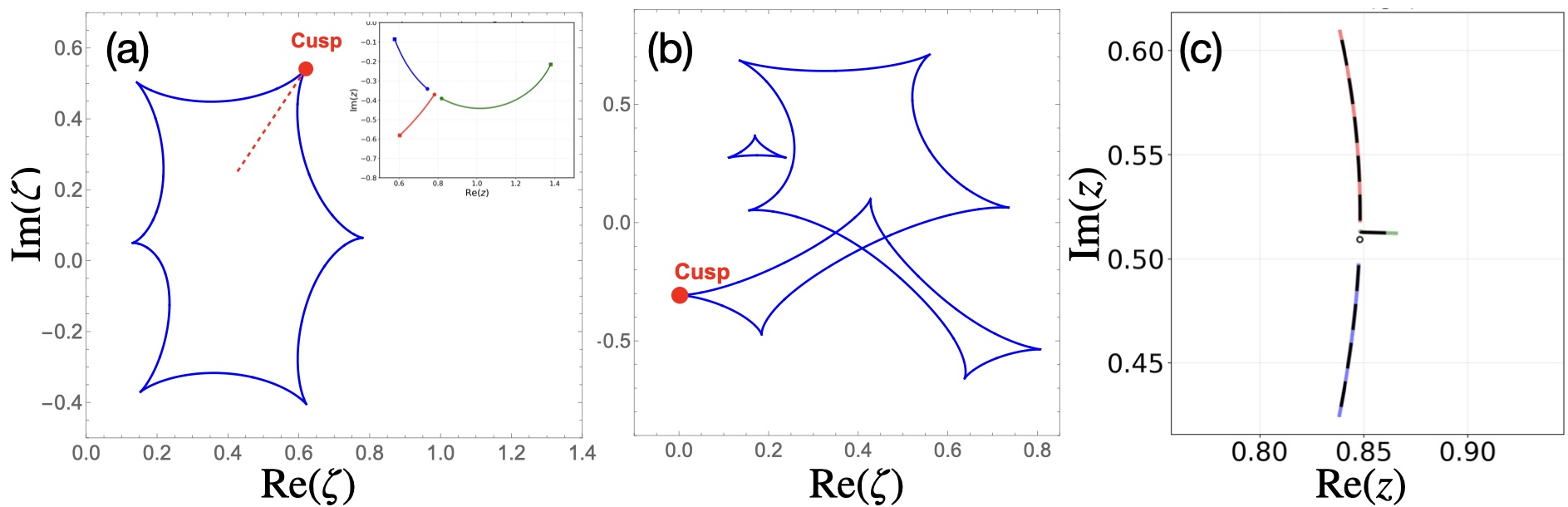}
\caption{\textbf{Caustic structure and image trajectories for two triple–lens configurations.}
\textbf{(a,b)} Source–plane caustics for a triple lens with mass fractions
$\varepsilon=(\tfrac12,\tfrac{3}{10},\tfrac15)$ at positions
$s=(0,1,1+3i)$ \textbf{(a)} and $s=(0,1,\tfrac{1+i}{2})$ \textbf{(b)}.
The analysed cusp is marked by the red dot; the red dashed segment indicates the initial source
direction (too short to be visible in \textbf{(b)}).
\textbf{Inset in (a):} image trajectories (Im\,$z$ vs.\ Re\,$z$) for the short track
$\zeta=\zeta_{*}-(2+3i)u$, $u\!\in\![0.001,0.1]$, reconstructed by the local geode construction:
evaluate the HC series $m(U)$ from the preparation–invariant kernel $\varphi$,
lift per branch $t=\omega\,m^{1/3}$ (with anisotropic seeds), and apply a short Newton polish
on the full eliminated decic $P(z;\zeta, \bar\zeta)=0$. Coloured solid curves: geode+Newton; black dashed curves:
exact roots of $P(z;\zeta,\bar\zeta)=0$.
\textbf{(c)} Near–tangential cusp passage for the configuration in \textbf{(b)}:
image trajectories as the source moves along the real axis from $u=5\times10^{-4}$ to $0.033$.
The same HC geode $\rightarrow$ lift $\rightarrow$ Newton construction recovers all three branches with
machine–precision residuals; the apparent “jump’’ in one branch is a projection effect from crossing
a nearby fold: continuation in the geode variable remains single-valued and smooth (Sec.~\ref{sec:triple-cusp}).
}
\label{fig:triple_cusp_trajectories}
\end{figure}

\section{Triple -- lens cusp (physical decic, local cubic)}
\label{sec:triple-cusp}

Our final examples apply the geode/HC construction to a fully physical
three–point–mass lens, where the global lens mapping is non–holomorphic in $z$
but the local imaging polynomial near a cusp is still captured by the same
preparation–invariant construction.  We consider two variants of the triple-lens cusp: a straightforward application of the geode method and a configuration that is geometrically
extreme: the source trajectory grazes the caustic almost tangentially, making
it one of the hardest cases for traditional root-tracking schemes. We treat two variants with the same masses $\varepsilon=(\tfrac12,\tfrac3{10},\tfrac15)$
but different positions $s=(0,1,1+3i)$ and $s=(0,1,(1+i)/2)$; Figures~\ref{fig:triple_cusp_trajectories}(a,b) mark
the analysed cusps (red dots).

We consider the Einstein-scaled three-point-mass lens
\begin{equation}
\zeta \;=\; z-\sum_{j=1}^{3}\frac{\varepsilon_j}{\overline z-\overline s_j}.
\label{eq:triple-lens-eq}
\end{equation}
The corresponding caustic networks are plotted in
Figures~\ref{fig:triple_cusp_trajectories}(a-b).

First, we will analyze the triple--lens cusp  with   $(s_1,s_2,s_3)=(0,\,1,\,1 + 3 i)$ (see Fig.~\ref{fig:triple_cusp_trajectories}(a)). The cusp (red dot in Fig.~\ref{fig:triple_cusp_trajectories}(a)) is given by
\begin{equation}
z_*=0.7803934 - 
 0.3667703 \,i,\qquad
\zeta_*=0.6199736 + 
 0.5411154\,i.
\label{eq:cusp_triple_lens_a}
\end{equation}
Section D of the Supplementary Information gives the details of the construction of the geode and HC series that follow the method we described in Sec.~\ref{sec:geode}. Here we only provide the outline of the construction. 

After local normalization and shift we consider the normalised local imaging polynomial 
\begin{equation}
P_{\rm loc}(w;v,\bar v)
=
P\!\bigl(z_{*}+w/\lambda;\,\zeta_{*}+v,\,\bar\zeta_{*}+\bar v\bigr),
\label{eq:triple-Ploc}
\end{equation}
centred at $(w,v,\bar v)=(0,0,0)$  with \(\partial_w^3 P_{\mathrm{loc}}(0;0,0)/3! = 1\) (ensured by \(\lambda = 3.179048207371207 - 0.970805399048372\,i\)).
  The Weierstrass centre $w_{c}(v,\bar v)$  defined by
$
\partial_{w}^{2}P_{\rm loc}\!\bigl(w_{c}(v,\bar v);v,\bar v\bigr)=0,$ $
w_{c}(0,0)=0,
$
is given by  
\begin{equation}
    w_c(v, \bar v) = (-0.6830560 - 1.404901\, i) v -(4.216304 + 2.530664 \, i) \bar{v} + O(\|(v,\bar v)\|^{2}).
\end{equation}
The preparation--invariant Lagrange form is $m=U\varphi(m)$ where
\begin{equation}
\varphi(m)=(1 - \alpha_1 m - \alpha_2 m^2 - \alpha_3 m^3 - \alpha_4 m^4)^{-1},
   \end{equation}
with $\alpha_1 = -(0.130049 + 0.105168\,i)$, $\alpha_2 = (0.00357267 - 0.00154033\, i),$ $\alpha_3=0.0000814471 + 0.0000559569\, i$ and $\alpha_4=1.21561\cdot 10^{-6},$ 
where we neglected the terms below $10^{-6}$. In the HC language of
Section~\ref{sec:HC-classification} this example has signature
$
\mathrm{Sig}_{R}=(3;\alpha_{1},\alpha_{2},\alpha_{3},\alpha_{4}).$

The inset of Fig.~\ref{fig:triple_cusp_trajectories}(a) shows three tracks emanating from the cusp along the source trajectory given by $\zeta=\zeta_* - (2 + 3\, i) u$. For $u=0.01$, we get $U=-0.00125007 - 0.00388306\,i$ and $m(u)=-0.00124729 - 0.0038829\, i$ from HC series. The lifted roots are $t_k=(0.310598 -0.477264\,i, -0.406215+0.193705 \, i,0.0956164 +0.283559\, i)$ and six steps of Newton polish bring the solutions to seven significant digits as $z_{1,2,3}=(0.9242055-0.4303363\,i ,0.6873154-0.2815570\, i,0.7720886-0.3837485\, i).$

For our second variety of the triple lens we consider $(s_1,s_2,s_3)=\Bigl(0,\,1,\,\tfrac{1+i}{2}\Bigr) $ with the cusp at 
\begin{equation}
z_*=0.8480548+0.5093864\,i,\qquad
\zeta_*=0.001904246-0.3071656\,i,
\label{eq:cusp_triple_lens}
\end{equation}
with cubic scale (pure $w$–jet monic)
$$
\lambda=\Bigl[\partial_{z}^{3}P(z_{*};\zeta_{*},\bar\zeta_{*})/3!\Bigr]^{1/3}
=0.3175679+0.2106751\,i.
$$
The normalised local imaging polynomial is again
\begin{equation}
P_{\rm loc}(w;v,\bar v)
=
P\!\bigl(z_{*}+(t + w_c)/\lambda;\,\zeta_{*}+v,\,\bar\zeta_{*}+\bar v\bigr) = \mathcal U(t;v,\bar v)\,W(t;v,\bar v).
\label{eq:triple-Ploc}
\end{equation}
Form $R_W(m,v,\bar v)=\prod_{j=0}^3W(\omega^j m^{1/3},v,\bar v)$ and observe that $R_W=
(-2.66759\cdot 10^{-7} - 3.9926 \cdot10^{-6}\, i) m^2 + (1.14296 + 0.331717\, i) m^3,$ so both constant and linear terms in $m$ are missing. This is the case of accidental multiplicity.  
 For the centred, monic depressed cubic
\(
W(t)=t^3+p\,t+q
\)
at a cusp (\([t^2]W\equiv0\)), the \(d=3\) cyclotomic product over the geode lifts \(t=\omega^j m^{1/3}\) has the closed form
\[
R_W(m; v,\bar v)
=\prod_{j=0}^{2}W\!\bigl(\omega^j m^{1/3}\bigr)
=\bigl(m+q\bigr)^3+p^{\,3}m
= q^3+\bigl(3q^2+p^{\,3}\bigr)m+3q\,m^2+m^3.
\]
Thus at parameter values where the combinations \(q\approx0\) and \(3q^2+p^{\,3}\approx0\) (numerically) hold, the constant and \(m^1\) terms are suppressed and one observes
$
R_W(m;v,\bar v)\approx a\,m^2+b\,m^3
$
(up to the nonvanishing unit from the Weierstrass factorisation). This  signals an \emph{accidental extra multiplicity in \(m\)} at that epoch. Consistently with Section~\ref{sec:geode}, we first form the square–free scalar $\widetilde R_W=R_W/\gcd_m(R_W,\partial_m R_W)$ and only then set  \(A=\partial_m \widetilde R_W(0)\), \(B=-\widetilde R_W(0)\), and the prepared source \(U=B/A\). In practice, dividing out the common \(m\)-factor (often \(m\) or \(m^2\) at such points) restores a linear-in-\(m\) scalar and a well-conditioned \(U\).  Dividing out $m^2$
 yields a linear-in-
$m$ scalar; solving it at this epoch gives
$m = 1.15032\times 10^{-6} + 3.15936\times 10^{-6} i$. We
proceed with the root lifting using the anisotropic seeds of equation~(\ref{eq:anisotropic-seed}).
 We get $t_k=(0.00792713 +0.00110449\, i,-0.0132836+0.0164089\, i,0.0053565 -0.0175134\, i)$ and after five steps of  Newton polish get the physical roots to the machine precision: $z_{1,2,3}=(0.8531663+0.5126516 \, i,0.84633898+0.5607243\, i,0.8440857+0.4603394\, i).$

\paragraph{ Preparation–invariant kernel \(\varphi(m)\).}
From the centred pure–$m$ jet of the square–free scalar,
\[
\widetilde R(m;0,0)=r_{3}m^{3}\Bigl(1+\delta_{1}m+\delta_{2}m^{2}+\delta_{3}m^{3}
+\delta_{4}m^{4}+\cdots\Bigr),
\]
we absorb \(r_3\) and define the preparation–invariant kernel
\[
\varphi(m) = \frac{1}{1 + \delta_1 m + \delta_2 m^2 + \delta_3 m^3 + \cdots}
= \Bigl(1 - \sum_{r\ge 1} \alpha_r m^r\Bigr)^{-1},
\]
with \(\alpha_r = -\delta_r\).
For this triple–lens cusp the first kernel coefficients are
\[
\alpha_{1}=-(249.579 + 360.576\,i),\quad
\alpha_{2}=135853. + 102306\,i,\quad
\alpha_{3}=-(1.63242\cdot10^7 + 9.44856 \cdot 10^6 \,i),
\]
with higher $\alpha_{r\le 10}$ nonzero as well.  In the HC language of
Section~\ref{sec:HC-classification} this example has signature
$
\mathrm{Sig}_{R}=(3;\alpha_{1},\alpha_{2},\alpha_{3},\alpha_{4},\ldots\alpha_{10}),
$
which is substantially “stiffer” (larger, more complex coefficients) than in
the previous triple lens example, reflecting the strongly distorted local geometry of this
 lens.

\paragraph{Certified radius and stiffness of the kernel.}
Using the majorant kernel
$\widehat\varphi(m)=(1-\sum_{r=1}^{4}|\alpha_{r}|m^{r})^{-1}$ and the
characteristic condition
$\sum_{r=1}^{4}(r+1)a_{r}m^{r}=1$ with $a_{r}=|\alpha_{r}|$, one obtains an indicative
 seed radius $\widehat\rho_{U}\approx 2.5\times 10^{-2}$ for the
HC series $m(U)$.  For the tangential track considered here we continue
slightly beyond this value (to $u\simeq 0.033$, corresponding to
$|U|\sim 0.03$) and observe that the geode+Newton reconstruction remains
accurate and stable, suggesting that the majorant–based $\widehat\rho_{U}$ is
 conservative.  The large, complex coefficients $\alpha_{r}$ thus
manifest as a ``stiff'' kernel with a relatively small guaranteed radius, but
the practical convergence extends somewhat further.

\paragraph{Geometric interpretation: near-tangential cusp crossing.}
The second triple-lens configuration considered here is geometrically exceptional: the
source trajectory, launched from the cusp point $\zeta_{*}$ along the real
axis, proceeds almost exactly along the caustic’s tangent.  The local caustic
branch at this cusp is inclined by barely $0.6^{\circ}$to the real axis, so as
the source advances along the real axis it does not pierce the caustic
frontally but slides along it: a near-glancing encounter that keeps the source
inside the cusp envelope over an extended interval of ${\rm Re}(\zeta)$.  A
displacement of order $10^{-4}$ in ${\rm Re}(\zeta)$ is sufficient to toggle the
system between four and six physical images, while a merging pair coalesces on
the critical curve before separating again.

Figure~\ref{fig:triple_cusp_trajectories}(c) displays the three image
trajectories in the complex image plane as the source moves along this
near–tangential track.  One of the branches (shown in green) exhibits a sharp
apparent ``jump'' across the critical curve.  This feature is not a numerical
artefact: a small change in ${\rm Re}(\zeta)$ forces the associated root to switch
sheets on the Riemann surface of the imaging polynomial, so its projection in
the $({\rm Re} (z),{\rm Im}(z))$–plane appears to jump, even though in the geode variable
$m$ the continuation remains perfectly analytic.  In the $m$–representation the
analytic branch passes smoothly through the fold where the images coalesce, and
the HC series of $m(U)$ remains valid throughout.  The discontinuity visible in
the plot is therefore imposed purely by the near-tangential geometry, while the
underlying geode expansion remains continuous and well-behaved.

This triple-lens cusp examples show that the geode/HC construction applies unchanged to
fully physical, non-holomorphic lens mappings and remains robust even in
geometrically extreme situations.  The preparation-invariant kernel
$\varphi(m)=\bigl(1-\sum_{r\ge1}\alpha_{r}m^{r}\bigr)^{-1}$ has large, complex
coefficients, leading to a relatively small certified seed radius
$\widehat\rho_{U}$, yet the HC seeds plus Newton on decic $P$ recover
all three image branches with machine–precision residuals along a
quasi-tangential source track.  In the geode variable the continuation across
the fold is smooth and single–valued; the apparent ``jump'' in
Figure~\ref{fig:triple_cusp_trajectories}(c) is entirely due to the projection
of the Riemann surface onto the $({\rm Re}(z), {\rm Im}(z))$-plane.  Together with
Examples~\ref{sec:ex-fold}--\ref{sec:decic-resonant}, this underscores the
central novelty of the method: a single, preparation–invariant analytic branch
suffices to track all local microlensing images across folds and cusps,
regardless of global degree, kernel complexity, or lens geometry.

\section{Consequences: Magnifications, Finite--Source Kernels, and Astrometric Centroids}
\label{sec:validation}

In this section we specialize the geode preparation of Section~\ref{sec:geode} to a fold ($d=2$) and record the resulting
preparation-invariant laws for point- and finite-source observables. All inputs are local jets at the multiple image; no global
solution of the lens polynomial is required. This recovers the standard fold scalings and finite–source kernels of
\citet{gaudi2002gravitationalI,gaudi2002gravitationalII,albrow2001planet} in a form where all lens dependence is compressed
into a small set of jet–derived invariants with certified error control. A consolidated comparison of HC 
signature and spectrum across our worked examples is deferred to Section~\ref{sec:HC-classification}.

\paragraph{(i) Fold normal form and geode.}
After centring and normalising (Section~\ref{sec:geode}), the local imaging
polynomial factors as
\[
P_{\rm loc}\bigl(w_{c}+t;v\bigr)
=
\mathcal U(t,v)\,W(t,v),\qquad
W(t,v)=t^{2}+p_{0}(v),\qquad \mathcal U(0,0)\neq0,
\]
where $v=\zeta-\zeta_{*}$ and $t=w-w_{c}(v)$, with $w_{c}(0)=0$ and
$p_{0}(0)=0$.  Introducing the fold geode $m=t^{2}$, the square–free scalar is given by (\ref{eq:2d})
so the image–side kernel is trivial, $\varphi\equiv 1$, and $m=-p_{0}(v)$ to
all orders.  In particular, the Lagrange variable $U$ coincides with $m$, and
the HC series collapses to $m(U)=U$ exactly.

Let $\tau$ be the signed normal offset in the source plane (positive on the
two–image side), and write
\[
p_{0}(v)=-\alpha_{0}\,\tau+O(\tau^{2}),\qquad \alpha_{0}>0,
\]
where $\alpha_{0}$ is a preparation–invariant function of the local jets
(captured explicitly in Section~\ref{sec:geode}).  Define the fold scales
\[
u_{f}=\alpha_{0},\qquad \alpha=u_{f}^{-1/2}.
\]
The two local branches in the centred image coordinate are then
\[
t_{\pm}(\tau)=\pm\sqrt{m}
=\pm\sqrt{\alpha_{0}\,\tau}\,+\,O(\tau)
=\pm\frac{\sqrt{\tau}}{\alpha}\,+\,O(\tau),
\]
realising the canonical $t\sim \pm \tau^{1/2}$ fold behaviour.

\paragraph{(ii) Point–source pair: magnification and centroid.}
Let $\mu_{\pm}(\tau)$ be the signed magnifications of the two fold images, and
set
\[
\mu_{\rm pair}(\tau)=\mu_{+}(\tau)+\mu_{-}(\tau),\qquad
\Theta_{\rm pair}(\tau)=\frac{\mu_{+}z_{+}+\mu_{-}z_{-}}{\mu_{\rm pair}(\tau)}.
\]
If $\hat{\boldsymbol n}$ is the unit normal to the critical curve at the base image, a standard jet calculation (invariant under admissible coordinate changes) gives
\[
\mu_{\rm pair}(\tau)
=\frac{\sqrt{u_{f}}}{\sqrt{\tau}}\Bigl(1+O(\tau^{1/2})\Bigr)
=\frac{1}{\alpha\sqrt{\tau}}\Bigl(1+O(\tau^{1/2})\Bigr),
\]
\[
\Theta_{\rm pair}(\tau)-z_{c}(0)
=B_{f}\,\tau^{1/2}\,\hat{\boldsymbol n}+O(\tau),\qquad B_{f}>0,
\]
where $u_{f}$ and $B_{f}$ are preparation–invariant functions of the local
jets.  Thus the usual $1/\sqrt{\tau}$ magnification and $\tau^{1/2}$ centroid
approach near a fold follow directly from the geode normal form with no global
solve, and all lens dependence is encoded in the scalars $(u_{f},B_{f})$.  In
the limit of vanishing source size these expressions coincide with the
classical fold asymptotics derived in
\citet{gaudi2002gravitationalII,gaudi2002gravitationalI}.

\paragraph{(iii) Finite–source photometry and astrometry.}
For a circular source of angular radius $\rho_{*}$ centred at signed distance
$\tau$, define the dimensionless offset
\[
s=\frac{\alpha\,\tau}{\rho_{*}}
=\frac{\tau}{\rho_{*}\sqrt{u_{f}}}.
\]
With linear limb–darkening coefficient $\Gamma$ the fold–pair magnification
takes the universal 1D kernel form
\[
\mu^{\rm fs}_{\rm pair}(\tau;\rho_{*})
=\rho_{*}^{-1/2}\Bigl[(1-\Gamma)\,G_{0}(s)+\Gamma\,G_{1/2}(s)\Bigr],
\]
where $G_{0}$ and $G_{1/2}$ are the standard chord–integral kernels and
$G_{\nu}(s)\sim s^{-1/2}$ as $s\to+\infty$, reproducing the point–source law.
The flux–weighted centroid projected on $\hat{\boldsymbol n}$ is
\begin{equation}
\bigl(\Theta^{\rm fs}_{\rm pair}(\tau;\rho_{*})-z_{c}(0)\bigr)\!\cdot\!\hat{\boldsymbol n}
=\frac{\rho_{*}^{1/2}}{\alpha}\,
\frac{(1-\Gamma)\,K_{0}(s)+\Gamma\,K_{1/2}(s)}{(1-\Gamma)\,G_{0}(s)+\Gamma\,G_{1/2}(s)}\,B_{f},
\qquad
\frac{K_{\nu}(s)}{G_{\nu}(s)}\sim s^{1/2}\ \ \text{as}\ s\to+\infty,
\end{equation}
 These expressions reproduce the fold–crossing kernels used in
\citet{gaudi2002gravitationalI,gaudi2002gravitationalII,albrow2001planet}, but
now with explicit identification of $u_{f}$ and $B_{f}$ as preparation–invariant
jet combinations and with a built–in truncation–error control from the geode
construction.

\paragraph{(iv) Analyticity and truncation control.}
For a fold the Lagrange equation is $m=U$ \emph{globally}, so there is no
branch point in the $U$–plane; the HC kernel is trivial and $m(U)$ is entire.
Practical truncation is therefore governed solely by the source–side series of
$p_{0}(\zeta)$ and $z_{c}(\zeta)$.  If
\[
p_{0}(\zeta)=\sum_{n\ge1}c_{n}(\zeta-\zeta_{*})^{n}
\]
is truncated at order $N$ and used on $|\zeta-\zeta_{*}|\le r$, Cauchy’s
estimate gives
\[
\left|\,p_{0}(\zeta)-\sum_{n=1}^{N}c_{n}(\zeta-\zeta_{*})^{n}\right|
\le \frac{M_{p_{0}}(r)}{r-|\zeta-\zeta_{*}|}
\left(\frac{|\zeta-\zeta_{*}|}{r}\right)^{\!N+1},
\quad
M_{p_{0}}(r)=\max_{|\xi-\zeta_{*}|=r}|p_{0}(\xi)|.
\]
With $m=-p_{0}$ and $t_{\pm}=\pm\sqrt{m}$, perturbations propagate as
\[
|\delta t_{\pm}|\le \tfrac12\,|m|^{-1/2}\,|\delta m|
=\tfrac12\,(\alpha_{0}\,\tau)^{-1/2}\,|\delta m|,
\]
and one Newton update on $W=t^{2}+p_{0}$,
\[
t\ \leftarrow\ t-\frac{t^{2}+p_{0}(\zeta)}{2t},
\]
squares the seed error whenever $t\neq 0$ (or one can use the exact
$t=\pm\sqrt{-p_{0}}$ very near the axis).  This provides a simple, fully
constructive truncation–error bound for the fold case, with all constants
determined by local jets.

\paragraph{(v) Cusps.}
For cusps ($d=3$) the same construction applies with $m=t^{3}$ and a nontrivial
preparation–invariant kernel $\varphi$ (Section~\ref{sec:geode}).  The HC
series $m(U)=\sum_{n\ge1}M_{n}U^{n}$ then has a finite analyticity radius
$\rho_{U}$ determined by the characteristic system; seeds
$t=\omega\,m^{1/3}$ are polished on the centred cubic $W$, as implemented in
Examples~\ref{sec:binary-cusp} and~\ref{sec:triple-cusp}.  The fold case
provides the simplest instance in which all these ideas can be written in
closed form while remaining directly relevant to finite–source photometry and
astrometric centroids.

\section{Classification by Hyper--Catalan signature and spectrum}
\label{sec:HC-classification}

The universal Lagrange form of Section~\ref{sec:geode} gives a
preparation–invariant description of every local multiple image:
\begin{equation}
m = U\,\varphi(m),\qquad \varphi(0)=1,\qquad U=\Lambda(\zeta,\bar \zeta),
\label{eq:summary}
\end{equation}
valid for any analytic lens polynomial and multiplicity $d\ge 2$.  This allows
the characteristics of optical catastrophes (folds, cusps, higher) to be
extended and quantified beyond their Thom–Arnold topological type.

In addition to the classical type (fold $A_{2}$, cusp $A_{3}$, etc.), we
introduce two complementary analytic moduli:
 the Hyper--Catalan (HC) signature, capturing intrinsic kernel
coefficients that control sparsity and stiffness of the single–series solution
$m(U)$ and 
 the HC spectrum, encoding the certified analyticity radius of
$m(U)$ via the location of branch points in the $U$–plane.
Together they quantify how “stiff’’ or “soft’’ a local mapping is and how fast
its HC expansion converges, features not captured by purely topological
classification.

\paragraph{Definition (HC signature).}
At the base point, write $\varphi(m)=(1-\sum_{r\ge1}\alpha_r m^r)^{-1}$.
The tuple $\mathrm{Sig}_R=(d;\alpha_1,\ldots,\alpha_R)$ of first nonzero $\alpha_r$
is the HC signature.

\paragraph{Definition (HC spectrum and radius).}
Branch points satisfy $\Delta(m)=\varphi(m)-m\varphi'(m)=0$ with 
$U=m/\varphi(m)$. The HC spectrum is $\Sigma=\{(m_\ast,U_\ast)\}$ and
$\rho_U=\min_{(m_\ast,U_\ast)\in\Sigma}|U_\ast|$.

In practice $R$ is small: for the binary–lens cusp (Example~\ref{sec:binary-cusp})
only $\alpha_{1},\alpha_{2}$ are nonzero, whereas the decic with resonant unit
(Example~\ref{sec:decic-resonant}) has a single nonzero $\alpha_{7}$ and the
triple–lens cusp (Example~\ref{sec:triple-cusp}) exhibits a “stiff’’ kernel
with several large complex $\alpha_{r}$. Beyond the Thom–Arnold topological type, observational analyses of microlensing variability demonstrate that subtle, quantitative features of the local caustic geometry carry physical information \citep{gil2005limits}. The HC signature isolates these analytic moduli in a preparation–invariant way.

\paragraph{Classification.}
For any local multiple image of multiplicity $d\ge 2$:
\begin{enumerate}\itemsep1pt
\item[(i)] Compute $w_{c}(v,\bar v)$ and the monic factor $W(t;v,\bar v)$
from the Weierstrass preparation of $P_{\rm loc}$.
\item[(ii)] Form the geode $m=t^{d}$ and the square–free scalar
$\widetilde R(m;v,\bar v)$ by cyclotomic product and gcd.
\item[(iii)] Extract the kernel $\varphi(m)$ from $\widetilde R(m;0,0)$ and
record the HC signature $\mathrm{Sig}_{R}=(d;\alpha_{1},\ldots,\alpha_{R})$.
\item[(iv)] Solve the characteristic system (\ref{eq:char}-\ref{eq:rhoU}) to obtain the HC spectrum $\Sigma$ and the analyticity
radius $\rho_{U}$.
\end{enumerate}
This classification separates topological type (fold versus cusp) from analytic
moduli (HC signature and spectrum) and yields a unified, preparation-invariant
description of all local microlensing caustics.

\section{Conclusion}
\label{sec:conclusion}

This work goes beyond the classical catastrophe description of folds and cusps
by introducing and applying, for the first time in lensing, the
preparation–invariant Hyper--Catalan (HC) construction.  The universal Lagrange
form (\ref{eq:summary})
is shown to govern \emph{all} local multiple images (folds, cusps, higher) of
analytic lens polynomials, with $U$ capturing source transport and
$\varphi$ capturing image–side geometry.  From this single-series framework we
obtain quantitative, model-agnostic moduli: the HC signature
$(d;\alpha_{1},\alpha_{2},\ldots)$ and the HC spectrum $\Sigma$, which provide
explicit sparsity structure for the series, certified convergence, and constructive tail bounds.
The construction is degree–agnostic and extends seamlessly to non-holomorphic
global maps (e.g. the triple lens) once locally Weierstrass-centred.

\paragraph{Decoupling and portability.}
The method separates source–side transport $U=\Lambda(v,\bar v)$ from the
image–side kernel $\varphi(m)$, so the same kernel governs different lenses
once they are centred, while $\Lambda$ carries model–specific geometry.  This
decoupling allows local kernels to be reused across models, with only the
source transport recomputed.
\paragraph{Certified reconstruction.}
The HC series $m(U)=\sum_{n\ge1}M_{n}U^{n}$ supplies analytic seeds
$t=\omega\,m^{1/d}$ ($\omega^{d}=1$) that reach machine precision after a short
Newton polish on the centred Weierstrass factor $W(t;v,\bar v)=0$, without
solving the global eliminant.  Majorant kernels $\widehat\varphi$ give explicit
truncation errors for any $|U|<\rho_{U}$, providing rigorous control of
approximation errors.
\paragraph{Consistency across models.}
The same construction (centre $\to$ monic $W$ $\to$ geode $m$ $\to$
$m=U\varphi(m)$) was validated on: (i) an artificial fold (recovering the
classical finite-source kernels and centroids with jet-based invariants),
(ii) a physical binary-lens cusp, (iii) an artificial decic cusp with a
nontrivial resonant kernel, and (iv)  physical triple-lens cusps with a
non–holomorphic global map.  In each case, the HC seeds plus Newton polish
reproduce all image branches with high precision inside the certified radius. The resulting analytic derivatives facilitate gradient-based inference, a capability increasingly required in modern microlensing constructions \citep{penny2019predictions, kains2009systematic}.
Such features enable real-time, robust analysis of fold/cusp crossings in surveys like OGLE or Roman.
\paragraph{Strict vs preparation–invariant gauge.}
In a strict gauge, point-mass folds/cusps have $\varphi_{\rm strict}\equiv 1$;
in the preparation-invariant convention adopted here, the base-point $m$-unit
is retained in $\varphi$, explaining the nontrivial kernels observed in the
binary and triple examples.  This choice
keeps all physically relevant resonances and analytic structure visible in the
HC signature.
\paragraph{Analytic derivatives and inference.}
Because $U$, $\varphi$, and $m(U)$ are analytic, all parameter derivatives
exist and are stable, enabling gradient-based inference with provable
convergence control.  Derivatives of image positions, magnifications, and
centroids with respect to lens parameters follow from termwise differentiation
of $m(U)$ and the local lift, rather than from finite–difference approximations
on high-degree root solvers.

\paragraph{Scope and outlook.}
The framework does not depend on the polynomial degree and accommodates resonant analytic units
(shear, ellipticity, multi–plane couplings) via the HC signature.  Future work
includes higher-$d$ singularities, certified multi-source compositions, and
data-driven priors on $\Lambda$ and $\{\alpha_{r}\}$ for inference in complex
lensing scenarios, especially in the regime of dense, Roman-like bulge monitoring campaigns \citep{Penny2019WFIRST}. Complex multi-year events like Gaia16aye, where a full Keplerian orbit and space parallax are required to fit the light curve \citep{Wyrzykowski2020Gaia16aye}, are natural targets for extensions of the geode formalism to fully dynamic lenses.

\paragraph{HC spectrum and signature of the worked examples.}
The  examples developed in
Sections~\ref{sec:ex-fold}--\ref{sec:triple-cusp} provide explicit instances
of the HC classification framework.  Example~\ref{sec:ex-fold} (artificial
fold) has signature $\mathrm{Sig}_{R}=(2;0)$ and trivial kernel
$\varphi\equiv 1$, so the geode $m(U)=U$ is entire and the HC spectrum is
empty, as expected for a fold.  Example~\ref{sec:binary-cusp} (binary–lens
cusp) has $\mathrm{Sig}_{R}=(3;\alpha_{1},\alpha_{2})$ with two small nonzero
coefficients and a moderate analyticity radius
$\rho_{U}\approx 0.56$, reflecting a relatively “soft’’ cusp of nearly
polynomial shape.  Example~\ref{sec:decic-resonant} (decic cusp with resonant
unit) is governed by a single nonzero coefficient at order seven, yielding the
sparse HC series with $\mathrm{Sig}_{R}=(3;\alpha_7)$ and a  radius
$\rho_{U}=0.651$, characteristic of a highly symmetric resonant kernel.  Finally,
Example~\ref{sec:triple-cusp} (triple–lens cusp) exhibits $
\mathrm{Sig}_{R}=(3;\alpha_{1},\alpha_{2},\alpha_{3},\alpha_{4})$ for a ``soft'' kernel variant and a “stiff’’ kernel  $
\mathrm{Sig}_{R}=(3;\alpha_{1},\cdots, \alpha_{10})$  with large coefficients $\alpha_r$ that grow in magnitude with $r$ and a small certified radius $\rho^{\mathrm{b}}_{U} \simeq 2.5\times10^{-2}$, quantifying
the extreme local geometry of the near–tangential cusp crossing.  Together
these fingerprints illustrate how the HC signature and spectrum provide a
unified, preparation–invariant set of analytic moduli for all local
microlensing singularities: folds, cusps, resonant units, and highly distorted
triple-lens caustics, while retaining practical predictive power for photometric
and astrometric observables.

The wider context extends beyond Galactic microlensing.  In quasar microlensing
and strong–lensing substructure, local fold–cusp patches dominate short–scale
behaviour; the same preparation/HC construction provides controllable surrogates
for image positions, magnifications, and time delays with certified errors \citep{belokurov2005point,belokurov2003light,evans2002microlensing,belokurov2002astrometric}.
For planetary statistics, OGLE’s 20-year census of wide-orbit planets \citep{poleski2021wide} illustrates the scale of data sets to which preparation–invariant caustic approaches could be applied. In
wave–optics regimes the HC spectrum sets the domain of uniform approximations,
while the germ $m(U)$ delivers systematic expansions of parameters entering
canonical diffraction integrals.  Plasma lensing and scintillation admit the
same local jets and therefore the same reduction.

There are natural limitations.  The construction is local and its efficacy is
bounded by $|U|<\rho_{U}$; approaching $\rho_{U}$ one must re-centre or pass to
uniform approximations informed by the multiplicity pattern in $\Sigma$.  When
caustics overlap or higher–codimension catastrophes are present, a fresh
preparation with the appropriate multiplicity $d$ is required.  The quality of
the invariants depends on accurate jet extraction; this is algorithmic but must
be implemented with care in models with time dependence, external shear, or
multi–plane structure.

Several extensions are immediate.  Automatic detection of multiplicity and jet
extraction can turn the preparation into a subroutine for inference
constructions, returning $\mathrm{Sig}_{R}$, $\Sigma$, $\rho_{U}$, and kernel
parameters together with certified error bounds.  Pre-tabulation of fold
kernels, centroid ratios, and HC coefficients can accelerate real–time
modelling in high-cadence surveys and astrometric monitoring.  On the theory
side, invariant formulas for higher HC coefficients and systematic treatment of
higher-codimension catastrophes and multi-plane maps follow from the same
elimination and Lagrange framework \citep{wildberger2025hyper,Gessel2016Lagrange,stanton1988recent,weierstrass1879einige,artin1998galois,Krantz2001Function}.  In all cases the central mechanism
remains the same: reduce to the univariate, preparation-invariant germ
$m=U\,\varphi(m)$, compute its Hyper--Catalan series and spectrum, and lift
once to the physical images.

\begin{acknowledgments}
N.G.B. acknowledges the support from  the HORIZON EIC Pathfinder Challenges project HEISINGBERG (grant 101114978), the EPSRC UK Multidisciplinary Centre for Neuromorphic Computing (grant UKRI982) and  the Weizmann–UK Make Connection grant (grant 142568).
\end{acknowledgments}

\begin{contribution}
G.B. came up with the initial research concept in the astrophysical context, performed numerical calculations, performed formal analysis and validation with some help from N.G.B. N.G.B. came up with the mathematical framework and supervised the project. Both authors wrote the manuscript. 
\end{contribution}

\bibliography{literature}{}
\bibliographystyle{aasjournalv7}

\end{document}